\documentclass[journal]{IEEEtran}

\usepackage{multicol}
\usepackage{amsmath,amssymb,cite,multirow,subfigure}
\usepackage{graphicx}
\usepackage{url}
\usepackage[linesnumbered,ruled, vlined]{algorithm2e}
\usepackage{algorithmic}
\usepackage{makecell}

\usepackage{booktabs}
\usepackage{threeparttable}

\begin{document}
%
\title{A High Throughput List Decoder Architecture\\ for Polar Codes}

\author{Jun Lin~\IEEEmembership{Student~Member,~IEEE}, Chenrong Xiong and Zhiyuan Yan,~\IEEEmembership{Senior~Member,~IEEE}
\thanks{Part of the preliminary results were presented at the 2014 IEEE Workshop
  on Signal Processing Systems (SiPS 2014)~\cite{jun_sips} and the 2015 IEEE
International Conference on Acoustics, Speech, and Signal Processing (ICASSP 2015)~\cite{psu_icassp}.}}

\maketitle

\begin{abstract}
While long polar codes can achieve the capacity of arbitrary binary-input discrete memoryless channels when decoded by a low complexity successive cancelation (SC) algorithm, the error performance of the SC algorithm is inferior for polar codes with finite block lengths. The cyclic redundancy check (CRC) aided successive cancelation list (SCL) decoding algorithm has better error performance than the SC algorithm. However, current CRC aided SCL (CA-SCL) decoders still suffer from long decoding latency and limited throughput. In this paper, a reduced latency list decoding (RLLD) algorithm for polar codes is proposed. Our RLLD algorithm performs the list decoding on a binary tree, whose leaves correspond to the bits of a polar code. In existing SCL decoding algorithms, all the nodes in the tree are traversed and all possibilities of the information bits are considered. Instead, our RLLD algorithm visits much fewer nodes in the tree and considers fewer possibilities of the information bits. When configured properly, our RLLD algorithm significantly reduces the decoding latency and hence improves throughput, while introducing little performance degradation. Based on our RLLD algorithm, we also propose a high throughput list decoder architecture, which is suitable for larger block lengths due to its scalable partial sum computation unit. Our decoder architecture has been implemented for different block lengths and list sizes using the TSMC 90nm CMOS technology. The implementation results demonstrate that our decoders achieve significant latency reduction and area efficiency improvement compared with other list polar decoders in the literature.
\end{abstract}


\section{Introduction}
\label{sec:intro}

Polar codes~\cite{arikan} are a significant breakthrough in coding theory, since they can achieve the channel capacity of binary-input symmetric memoryless channels~\cite{arikan} and arbitrary discrete memoryless channels~\cite{sas_polar}. Polar codes of block length $N$ can be efficiently decoded by a successive cancelation (SC) algorithm~\cite{arikan} with a complexity of $O(N\log N)$. While polar codes of very large block length ($N>2^{20}$~\cite{gross_polar1}) approach the capacity of underlying channels under the SC algorithm, for short or moderate polar codes, the error performance of the SC algorithm is worse than turbo or LDPC codes~\cite{ido_it}.

Lots of efforts~\cite{ido_it, list2, list3} have already been devoted to the improvement of error performance of polar codes with short or moderate lengths. An SC list (SCL) decoding algorithm~\cite{ido_it} performs better than the SC algorithm. In~\cite{ido_it,list2, list3}, the cyclic redundancy check (CRC) is used to pick the output codeword from $L$ candidates, where $L$ is the list size. The CRC-aided SCL (CA-SCL) decoding algorithm performs much better than the SCL decoding algorithm at the expense of negligible loss in code rate.

Despite its significantly improved error performance, the hardware implementations of SC based list decoders~\cite{tree_list_dec, jun_low_mem_list, yuan_llr, llr_list_tsp, llr_list} still suffer from long decoding latency and limited throughput due to the serial decoding schedule.
In order to reduce the decoding latency of an SC based list decoder, $M$  $(M >1)$ bits are decoded in parallel in~\cite{bin_polar, yuan_low_latency, chenrong_tsp}, where the decoding speed can be improved by $M$ times ideally. However, for the hardware implementations of the algorithms in~\cite{bin_polar, yuan_low_latency, chenrong_tsp}, the actual decoding speed improvement is less than $M$ times due to extra decoding cycles on finding the $L$ most reliable paths among $2^ML$ candidates, where $L$ is list size.
A software adaptive SSC-list-CRC decoder was proposed in~\cite{gabi_low_latency}. For a (2048, 1723) polar+CRC-32 code, the SSC-list-CRC decoder with $L=32$ was shown to be about 7 times faster than an SC based list decoder. However, it is unclear whether the list decoder in~\cite{gabi_low_latency} is suitable for hardware implementation.

In this paper, a tree based reduced latency list decoding algorithm and its corresponding high throughput architecture are proposed for polar codes. The main contributions are:
\begin{itemize}
\item A tree based reduced latency list decoding (RLLD) algorithm over logarithm likelihood ratio (LLR) domain is proposed for polar codes. Inspired by the simplified successive cancelation (SSC)~\cite{low_latency_polar} decoding algorithm and the ML-SSC algorithm\cite{ml_ssc}, our RLLD algorithm performs the SC based list decoding on a binary tree. Previous SCL decoding algorithms visit all the nodes in the tree and consider all possibilities of the information bits, while our RLLD algorithm visits much fewer nodes in the tree and considers fewer possibilities of the information bits. When configured properly, our RLLD algorithm significantly reduces the decoding latency and hence improves throughput, while introducing little performance degradation.
\item Based on our RLLD algorithm, a high throughput list decoder architecture is proposed for polar codes. Compared with the state-of-the-art SCL decoders in~\cite{llr_list_tsp,jun_low_mem_list,yuan_low_latency}, our list decoder achieves lower decoding latency and higher area efficiency (throughput normalized by area).

\end{itemize}

More specifically, the major innovations of the proposed decoder architecture are:
\begin{itemize}
\item An index based partial sum computation (IPC) algorithm is proposed to avoid copying partial sums directly when one decoding path needs to be copied to another. Compared with the lazy copy algorithm in~\cite{ido_it}, our IPC algorithm is more hardware friendly since it copies only path indices, while the lazy copy algorithm needs more complex index computation.
\item Based on our IPC algorithm, a hybrid partial sum unit (Hyb-PSU) is proposed so that our list decoder is suitable for larger block lengths. The Hyb-PSU is able to store most of the partial sums in area efficient memories such as register file (RF) or SRAM, while the partial sum units (PSUs) in~\cite{llr_list_tsp, jun_low_mem_list,tree_list_dec} store partial sums in registers, which need much larger area when the block length $N$ is larger. Compared with the PSU of~\cite{jun_low_mem_list}, our Hyb-PSU achieves an area saving of 23\% and 63\% for block length $N=2^{13}$ and $2^{15}$, respectively, under the TSMC 90nm CMOS technology.
\item For our RLLD algorithm, when certain types of nodes are visited, each current decoding path splits into multiple ones, among which the $L$ most reliable paths are kept. In this paper, an efficient path pruning unit (PPU) is proposed to find the $L$ most reliable decoding paths among the split ones. For our high throughput list decoder architecture, the proposed PPU is the key to the implementation of our RLLD algorithm.
\item For the fixed-point implementation of our RLLD algorithm, a memory efficient quantization (MEQ) scheme is used to reduce the number of stored bits. Compared with the conventional quantization scheme, our MEQ scheme reduces the number of stored bits by 17\%, 25\% and 27\% for block length $N=2^{10}$, $2^{13}$ and $2^{15}$, respectively, at the cost of slight error performance degradation.
\end{itemize}

Note that the SSC and ML-SSC algorithms reduce the decoding latency by first performing it on a binary tree and then pruning the binary tree. Inspired by this idea, our RLLD algorithm performs the SC based list decoding algorithm on a binary tree. The low-latency list decoding algorithm~\cite{gabi_low_latency} also performs the list decoding algorithm on a binary tree. Our work~\cite{jun_sips} and the decoding algorithm in~\cite{gabi_low_latency} are developed independently. While both our RLLD algorithm and the low-latency list decoding algorithm in~\cite{gabi_low_latency} visit fewer nodes in the binary tree so as to reduce the decoding latency, there are some differences:
\begin{itemize}
\item Compared with the decoding algorithm in~\cite{gabi_low_latency}, our RLLD algorithm visits fewer nodes. Illuminated by the ML-SSC algorithm, our RLLD algorithm processes certain arbitrary rate nodes~\cite{low_latency_polar} in a fast way.
\item When a rate-1 node~\cite{low_latency_polar} is visited, our RLLD algorithm employs a less complex and hardware friendly algorithm to compute the returned constituent codewords.
\item Our RLLD algorithm is based on LLR messages, while the algorithm in~\cite{gabi_low_latency} is based on logarithm likelihood (LL) messages, which require a larger memory to store.
\end{itemize}

In terms of hardware implementations, compared with state-of-the-art SC list decoders~\cite{tree_list_dec, jun_low_mem_list, llr_list_tsp,llr_list, yuan_low_latency, chenrong_tsp}, our high throughput list decoder architecture shows advantages in various aspects:
\begin{itemize}
\item For the high throughput list decoder architecture, LLR message is employed while LL message was used in~\cite{tree_list_dec, jun_low_mem_list, yuan_low_latency, chenrong_tsp}. The LL based memories require more quantization bits and a larger memory to store. The area efficient memory architecture in~\cite{jun_low_mem_list} is employed to store all LLR messages. LLR messages were also employed in~\cite{llr_list_tsp,llr_list}. However, the register based memories in~\cite{llr_list_tsp, llr_list} suffer from excessive area and power consumption when $N$ is large.
\item Our list decoder architecture employs a Hyb-PSU, which is scalable for polar codes of large block lengths. The register based PSUs of the list decoders in~\cite{llr_list_tsp, jun_low_mem_list,tree_list_dec} suffer from area overhead when the block length is large. Instead of copying partial sums directly, our scalable PSU copies only decoding path indices, which avoids additional energy consumption.
\end{itemize}
The proposed high throughput list decoder architecture has been implemented for several block lengths and list sizes under the TSMC 90nm CMOS technology. The implementation results show that our decoders outperform existing SCL decoders in both decoding latency and area efficiency. For example, compared with the decoders of~\cite{llr_list_tsp}, the area efficiency and decoding latency of our decoders are 1.59 to 32.5 times and 3.4 to 6.8 times better, respectively.

For our RLLD algorithm and the corresponding decoder architecture, when computing the returned constituent codewords from an FP node or a rate-1 node, the returned $L$ constituent codewords may not be the $L$ most reliable ones among all candidates. This kind of approximation leads to more efficient hardware implementation of our list decoding algorithm at the cost of certain performance degradation. In contrast, existing SC list decoders in~\cite{ido_it,llr_list} usually selects the $L$ most reliable candidates.

The rest of the paper is organized as follows. Related preliminaries are reviewed in Section~\ref{sec: pre}. The proposed RLLD algorithm is presented in Section~\ref{sec: rlld}. The high throughput list decoder architecture is presented in Section~\ref{sec: llldec}. In Section~\ref{sec: imp_results}, the implementation and comparisons results are shown. At last, the conclusion is drawn in Section~\ref{sec: conclusion}.

\section{Preliminaries}
\label{sec: pre}

\subsection{Polar Codes} \label{ssec:polar_encoding}
\begin{figure} [hbt]
\centering
  \includegraphics[width=2.1in]{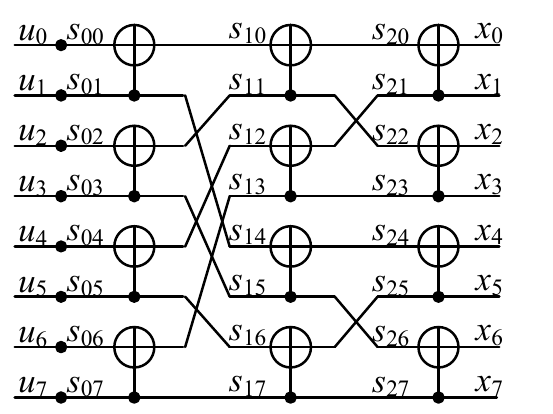}
  \caption{Polar encoder with $N=8$}\label{fig: encoding}
\end{figure}

Let $u_0^{N-1} = (u_0,u_1,\cdots,u_{N-1})$ denote the data bit sequence and $x_0^{N-1} = (x_0,x_1,\cdots,x_{N-1})$ the corresponding codeword, where $N=2^n$. Under the polar encoding, $x_0^{N-1} = u_0^{N-1}B_NF^{\otimes n}$ $(n\geqslant1)$, where $B_N$ is the bit reversal permutation matrix, and $F=\left[{1\atop 1}{0\atop 1}\right]$. Here $\otimes n$ denotes the $n$th Kronecker power, $F^{\otimes n} = F\otimes F^{\otimes (n-1)}$ and $F^{\otimes0}=1$.
For $i=0,1,\cdots,N-1$, $u_i$ is either an information bit or a frozen bit, which is set to zero usually. For an $(N,K)$ polar code, there are a total of $K$ information bits within $u_0^{N-1}$.
The encoding graph of a polar code with $N=8$ is shown in Fig.~\ref{fig: encoding}.

\subsection{Prior Tree-Based SC Algorithms} \label{ssec: ssc}

\begin{figure} [hbt]
\centering
  \includegraphics[width=2.8in]{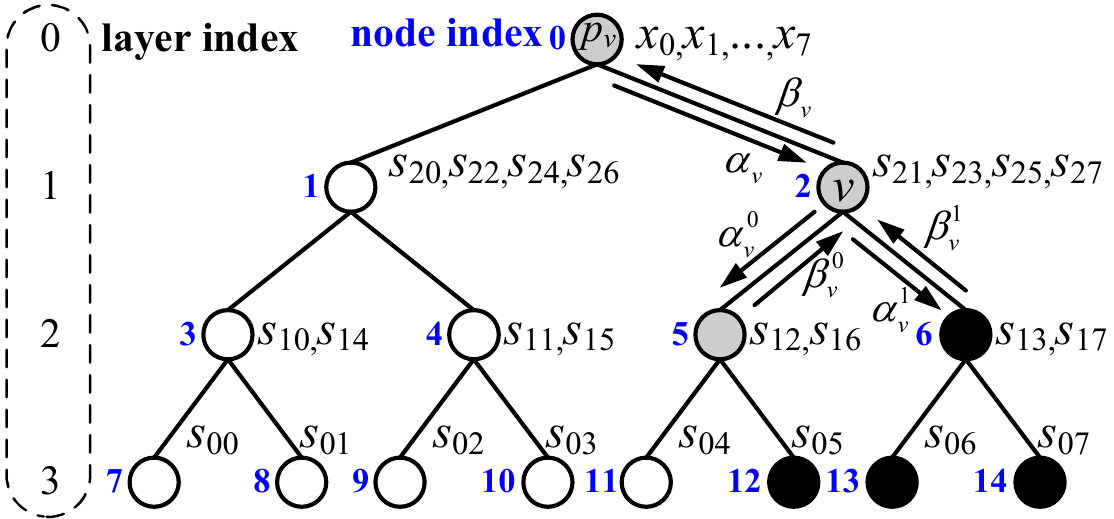}
  \caption{Binary tree representation of an (8, 3) polar code}\label{fig: dec_tree}
\end{figure}

A polar code of block length $N=2^n$ can also be represented by a full binary tree $G_n$ of depth $n$~\cite{low_latency_polar}, where each node of the tree is associated with a constituent code. For example, for node 1 shown in Fig.~\ref{fig: dec_tree}, the correspondent constituent code is the set $\{(s_{20},s_{22},s_{24},s_{26})\}$, where each element $(s_{20},s_{22},s_{24},s_{26})$ relates to the data word $u_0^{7}$ as shown in Fig.~\ref{fig: encoding}. The binary tree representation of an (8, 3) polar code is shown in Fig.~\ref{fig: dec_tree}, where the black and white leaf nodes correspond to information and frozen bits, respectively. 
There are three types of nodes in a binary tree representation of a polar code: rate-0 , rate-1 and arbitrary rate nodes. The leaf nodes of a rate-0 and rate-1 nodes correspond to only frozen and information bits, respectively. The leaf nodes of an arbitrary rate node are associated with both information and frozen bits. The rate-0, rate-1 and arbitrary rate nodes in Fig.~\ref{fig: dec_tree} are represented by circles in white, black and gray, respectively.

The SC algorithm can be mapped on $G_n$, where each node acts as a decoder for its constituent code. The SC algorithm is initialized by feeding the root node with the channel LLRs, ($Y_0, Y_1, \cdots, Y_{N-1}$), where $Y_i = \log(\Pr(y_i|x_i=0)/\Pr(y_i|x_i=1))$ and $(y_0,y_1,\cdots,y_{N-1})$ is the received channel message vector. As shown in Fig.~\ref{fig: dec_tree}, the decoder at node $v$ receives a soft information vector $\alpha_v$ and returns a constituent codeword $\beta_v$. When a non-leaf node $v$ is activated by receiving an LLR vector $\alpha_v$, it calculates a soft information vector $\alpha_{v}^0$ and sends it to its left child.
Node $v$ first waits until it receives a constituent codeword $\beta_v^0$, and then computes and sends a soft information vector $\alpha_v^1$ to its right child.
Once the right child returns a constituent codeword $\beta_v^1$, node $v$ computes and returns a constituent codeword $\beta_v$.
When a leaf node $v$ is activated, the returned constituent codeword $\beta_v$ contains only one bit $\beta_v[0]$, where $\beta_v[0]$ is set to 0 if leaf node $v$ is associated with a frozen bit; otherwise, $\beta_v[0]$ is calculated by making a hard decision on the received LLR $\alpha_v[0]$, where
\begin{eqnarray}\label{equ: hard_dec}
 \beta_v[0] = h(\alpha_v[0])= \left\{ \begin{array}{ll}
 0 & \alpha_v[0] \geqslant 0,\\
 1 & \alpha_v[0] < 0.
 \end{array} \right.
\end{eqnarray}
From the root node, all nodes in a tree are activated in a recursive way for the SC algorithm. Once $\beta_v$ for the last leaf node is generated, the codeword $x_0^{N-1}$ can be obtained by combining and propagating $\beta_v$ up to the root node.

The SSC decoding algorithm in~\cite{low_latency_polar} simplifies the processing of both rate-0 and rate-1 nodes. Once a rate-0 node is activated, it immediately returns the all zero vector. Once a rate-1 node is activated, a constituent codeword is directly calculated by making hard decisions on the received soft information vector as shown in Eq.~(\ref{equ: hard_dec}). The ML-SSC decoding algorithm~\cite{ml_ssc} further accelerates the SSC decoding algorithm by performing the exhaustive-search ML decoding on some resource constrained arbitrary rate nodes, which are called ML nodes in~\cite{ml_ssc}. For an ML node with layer index $t$, the constituent codeword passed to the parent node $p_v$ is
\begin{equation}\label{equ: mld}
\beta_v = \arg\!\max_{\textbf{x}\in \mathcal{C}}\sum_{i=0}^{2^{n-t}-1}(1-2\textbf{x}[i])\alpha_v[i],
\end{equation}
where $\mathcal{C}$ is the constituent code associated with node $v$.

\subsection{LLR Based List Decoding Algorithms} \label{ssec: llr_based}
For SCL decoding algorithms~\cite{ido_it, tree_list_dec, llr_list}, when decoding an information bit $u_i$, each decoding path splits into two paths with $\hat{u}_i$ being 0 and 1, respectively. Thus $2L$ path metrics are computed and the $L$ paths correspond to the $L$ minimum path metrics are kept.
The list decoding algorithms~\cite{ido_it, tree_list_dec} are performed either on probability or logarithmic likelihood (LL) domain. In~\cite{llr_list}, an LLR based list decoding algorithm was proposed to reduce the message memory requirement and the computational complexity of LL based list decoding algorithm. For decoding path $l$ $(l=0,1,\cdots,L-1)$, the LLR based list decoding algorithm employs a novel approximated path metric
\begin{equation}\label{equ: path_metric}
\mbox{PM}_l^{(i)} = \sum_{k=0}^{i}D(L_n^{(k)}[l],\hat{u}_k[l]),
\end{equation}
where $D(L_n^{(k)}[l],\hat{u}_k[l])$ is set to 0 if $h(L_n^{(k)}[l])$ equals $\hat{u}_k[l]$ or $|L_n^{(k)}[l]|$ otherwise. Here $L_n^{(k)}[l] \triangleq \log\frac{W_n^{(k)}(y_0^{N-1},\hat{u}_0^{k-1}[l]|0)}{W_n^{(k)}(y_0^{N-1},\hat{u}_0^{k-1}[l]|1)}$ and $y_0^{N-1}=(y_0,y_1,\cdots,y_{N-1})$ is the received channel message vector.
\section{Reduced Latency List Decoding Algorithm}
\label{sec: rlld}

\subsection{SCL Decoding on A Tree}\label{ssec:node_activation}
\begin{figure} [hbt]
\centering
  \includegraphics[width=2.6in]{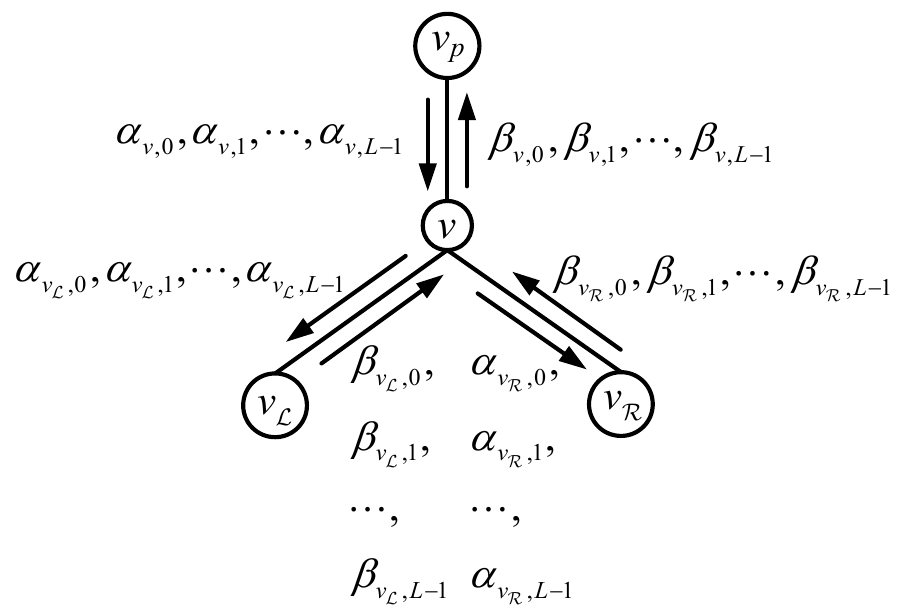}
  \caption{Node activation schedule for SC based list decoding on $G_n$}\label{fig: list_node_op}
\end{figure}

Similar to the SSC decoding algorithm, we also perform the SC based list decoding algorithms~\cite{ido_it, tree_list_dec} on a full binary tree $G_n$~\cite{jun_sips, gabi_low_latency}. The SCL decoding is initiated by sending the received channel LLR vector to the root node of $G_n$. As shown in Fig.~\ref{fig: list_node_op}, without losing generality, each internal node $v$ in $G_n$ is activated by receiving $L$ LLR vectors, $\alpha_{v,0}, \alpha_{v,1}, \cdots, \alpha_{v,L-1}$, from its parent node $v_p$ and is responsible for producing $L$ constituent codewords, $\beta_{v,0}, \beta_{v,1}, \cdots, \beta_{v,L-1}$, where $\alpha_{v,l}$ and $\beta_{v,l}$ correspond to decoding path $l$ for $l=0,1,\cdots,L-1$. Suppose the layer index of node $v$ is $t$, $\alpha_{v,l}$ and $\beta_{v,l}$ have $2^{n-t}$ LLR messages and binary bits, respectively, for $l=0,1,\cdots,L-1$.

Once a non-leaf node $v$ is activated, it calculates $L$ LLR vectors, $\alpha_{v_\mathcal{L},0}, \alpha_{v_\mathcal{L},1}, \cdots, \alpha_{v_\mathcal{L},L-1}$, and passes them to its left child node $v_\mathcal{L}$, where
\begin{equation}\label{equ: f_comp}
\alpha_{v_\mathcal{L},l}[i] = f(\alpha_{v,l}[2i],\alpha_{v,l}[2i+1])
\end{equation}
for $0\leq i <2^{n-t-1}$ and $l=0,1,\cdots,L-1$. Here $f(a,b) = 2\tanh^{-1}(\tanh(a/2)\tanh(b/2))$ and can be approximated as:
\begin{equation}\label{equ: f_comp_simplified}
f(a,b)\thickapprox \mbox{sign}(a)\cdot \mbox{sign}(b)\min(|a|,|b|).
\end{equation}

Node $v$ then waits until it receives $L$ codewords, $\beta_{v_\mathcal{L},0}$, $\beta_{v_\mathcal{L},1}$, $\cdots$, $\beta_{v_\mathcal{L},L-1}$, from $v_\mathcal{L}$. In the following step, node $v$ calculates another $L$ LLR vectors, $\alpha_{v_\mathcal{R},0}$, $\alpha_{v_\mathcal{R},1}$, $\cdots$, $\alpha_{v_\mathcal{R},L-1}$, and passes them to its right child node $v_\mathcal{R}$, where
\begin{eqnarray}\label{equ: g_comp}
 \begin{array}{lll}
 \alpha_{v_\mathcal{R},l}[i] &=& g(\alpha_{v,l}[2i], \alpha_{v,l}[2i+1], \beta_{v_\mathcal{L},l}[i])\\
  & = & \alpha_{v,l}[2i](1-2\beta_{v_\mathcal{L},l}[i])+\alpha_{v,l}[2i+1]
 \end{array}
\end{eqnarray}
for $0\leq i <2^{n-t-1}$ and $l=0,1,\cdots,L-1$.

At last, after node $v$ receives $L$ codewords, $\beta_{v_\mathcal{R},0}$, $\beta_{v_\mathcal{R},1}$, $\cdots$, $\beta_{v_\mathcal{R},L-1}$, from $v_\mathcal{R}$, it calculates $\beta_{v,0}$, $\beta_{v,1}$, $\cdots$, $\beta_{v,L-1}$ and passes them to its parent node $v_p$, where
\begin{equation}\label{equ: partial_sum}
(\beta_{v,l}[2i], \beta_{v,l}[2i+1]) = (\beta_{v_\mathcal{L}, l}[i]\oplus\beta_{v_\mathcal{R}, l}[i], \beta_{v_\mathcal{R}, l}[i]),
\end{equation}
for $0\leq i <2^{n-t-1}$ and $l=0,1,\cdots,L-1$.

For $l=0,1,\cdots,L-1$, PM$_l$ is the path metric associated with decoding path $l$ and is initialized with 0.
When a leaf node $v$ associated with an information bit is activated, decoding path $l$ splits into two paths with $\beta_{v,l}$ being 0 and 1, respectively. Note that the layer index of a leaf node is $n$, hence $\alpha_{v,l}$ and $\beta_{v,l}$ have only one LLR and binary bit, respectively, when node $v$ is a leaf node. For the SCL decoding, $2L$ expanded path metrics are computed, where
\begin{equation}\label{equ: path_metric}
\mbox{PM}_l^j = \mbox{PM}_l+D(\alpha_{v,l},j),
\end{equation}
for $j=0,1$ and $l=0,1,\cdots,L-1$. $D(\alpha_{v,l},j) = 0$ if $h(\alpha_{v,l})$ equals $j$. Otherwise, $D(\alpha_{v,l},j) = |\alpha_{v,l}|$. Suppose the $L$ minimum expanded path metrics are $\mbox{PM}_{a_0}^{j_0}$, $\mbox{PM}_{a_1}^{j_1}$, $\cdots$, $\mbox{PM}_{a_{L-1}}^{j_{L-1}}$, which correspond to the $L$ most reliable paths, then $\beta_{v,l}=j_l$ for $l=0,1,\cdots,L-1$. Decoding path $a_l$ will be copied to decoding path $l$ before further partial sum and LLR vector computations. For each decoding path $l$, path metric is also updated with $\mbox{PM}_l = \mbox{PM}_{a_l}^{j_l}$.
When a leaf node $v$ associated with a frozen bit is activated, $\beta_{v,l} =0$ for $l=0,1,\cdots,L-1$ are passed to its parent node $v_p$. The updated path metric PM$_l$ = PM$_l$ + $D(\alpha_{v,l},0)$.

The SCL algorithm on a tree described above is equivalent to the SCL algorithms in~\cite{ido_it, tree_list_dec}.


\subsection{Proposed RLLD algorithm}
In this paper, a reduced latency list decoding (RLLD) algorithm is proposed to reduce the decoding latency of SC list decoding for polar codes. For a node $v$, let $I_v$ denote the total number of leaf nodes that are associated with information bits. Let $X_{th}$ be a predefined threshold value and $X_0$ and $X_1$ be predefined parameters. Our RLLD algorithm performs the SC based list decoding on $G_n$ and follows the node activation schedule in Section~\ref{ssec:node_activation}, except when certain type of nodes are activated. These nodes calculate and return the codewords to their parent nodes while updating the decoding paths and their metrics, without activating their child nodes. Specifically:
\begin{itemize}
\item When a rate-0 node $v$ is activated, $\beta_{v,l}$ is a zero vector for $l=0,1,\cdots,L-1$.
\item When a rate-1 node $v$ with $I_v>X_{th}$ is activated, $\beta_{v,l}$ is just the hard decision of $\alpha_{v,l}$ for $l=0,1,\cdots,L-1$. For polar codes constructed in~\cite{arikan_code_construct, tal_code_construct}, we observe that the polarized channel capacities of the information bits corresponding to rate-1 nodes with $I_v> X_{th}$ are greater than those of the other information bits. Hence, for rate-1 nodes with $I_v> X_{th}$, our RLLD algorithm considers only the most reliable candidate codeword for each decoding path due to a more reliable channel.
\item When a rate-1 node $v$ with $I_v\leqslant X_{th}$ is activated, the returned codewords are calculated by a candidate generation (CG) algorithm, which is proposed later.
\item Let $t$ denote the layer index of node $v$. When an arbitrary rate node $v$ with $I_v\leqslant X_0$ and $2^{n-t} \leqslant X_1$ is activated, each decoding path splits into $2^{I_v}$ paths. From now on, such an arbitrary rate node is called fast processing (FP) node. A metric based search (MBS) algorithm, which is proposed later, is used to calculate the returned codewords.
\end{itemize}
Moreover, our RLLD algorithm works on a pruned tree. As a result, our RLLD algorithm visits fewer nodes than the SCL algorithm in~\cite{ido_it, tree_list_dec}. The full binary tree is pruned in the following ways:
\begin{itemize}
\item Starting from the complete tree representation of a polar code, label all FP nodes such that the parent node of each of them is not an FP node. Note that an FP node $v$ is an arbitrary rate node with $I_v\leqslant X_0$ and $2^{n-t} \leqslant X_1$. For each labeled FP node, remove all its child nodes.
\item Based on the pruned tree from the previous step, label all rate-0 and rate-1 nodes such that the parent node of each of these rate-0 and rate-1 nodes is not a rate-0 and rate-1 node, respectively. In the next, remove all child nodes of each of labeled rate-0 and rate-1 node.
\end{itemize}
The leaf nodes of the pruned tree from the above two steps consist of rate-0, rate-1 and FP nodes. The non-leaf nodes of the pruned tree are arbitrary rate nodes.

When a rate-1 node with $I_v > X_{th}$ or a rate-0 node is activated, ideally PM$_l$ is updated with PM$_l$ + $\Delta_{v,l}$ for $l=0,1,\cdots,L-1$, where $\Delta_{v,l} = \sum_{i=0}^{I_v-1}D(\alpha_{v,l}[i], \beta_{v,l}[i])$. For each rate-1 node with $I_v > X_{th}$, $\Delta_{v,l} =0$ since $\beta_{v,l}$ is the hard decision of $\alpha_{v,l}$. However, for a rate-0 node, $\Delta_{v,l}$ could have a non-zero value. For our RLLD algorithm, $\Delta_{v,l}$ is also set to 0 for each rate-0 node, since the resulting performance degradation is negligible. By setting $\Delta_{v,l}$ to 0, we no longer need to calculate $\alpha_{v,l}$ sent to a rate-0 node.

\subsubsection{Proposed CG Algorithm}

When a rate-1 node with $I_v \leqslant X_{th}$ is activated, instead of considering $2^{I_v}$ candidate codewords for each decoding path, since there are at most $L$ codewords from the same decoding path that could be passed to the parent node, it is enough to find only the $L$ most reliable codewords among $2^{I_v}$ candidates for each decoding path. When $I_v$ is large (e.g. $I_v\geqslant 32$), finding the $L$ most reliable codewords is computationally intensive and lacks efficient hardware implementations. For our RLLD algorithm, we considers only the $W (W<L)$ most reliable codewords among $2^{I_v}$ candidates for each decoding path. In this paper, $W$ is set to 2, since it results in efficient hardware implementations at the cost of negligible performance loss.

When $W=2$, the proposed CG algorithm, shown in Alg.~\ref{algo: CG}, is used to calculate the codewords passed to the parent node. Besides, the CG algorithm also outputs $L$ list indices, $a_0, a_1,\cdots,a_{L-1}$, which indicate that decoding path $a_l$ needs to be copied to path $l$. Suppose the layer index of such a rate-1 node $v$ is $t$. For each decoding path $l$, there are $2^{I_v} = 2^{2^{n-t}}$ candidate codewords that could be passed to the parent node $v_p$. However, our CG algorithm considers only the most reliable codeword $\mathcal{C}_{v,l,0}$ and the second most reliable codeword $\mathcal{C}_{v,l,1}$. In order to find these two codewords, each candidate codeword $\mathcal{C}_{v,l,j}$ is associated with a node metric
\begin{equation}\label{equ: direct_method}
\mbox{NM}_l^j = \sum\nolimits_{k=0}^{I_v-1}m_k|\alpha_{v,l}[k]|
\end{equation}
for $j=0,1,\cdots,2^{I_v}-1$, where $m_k=0$ if $\mathcal{C}_{v,l,j}[k]$ equals $h(\alpha_{v,l}[k])$ and 1 otherwise. As a result, the smaller a node metric is, the more reliable the corresponding candidate codeword is. Based on Eq.~(\ref{equ: direct_method}), $\mathcal{C}_{v,l,0} = h(\alpha_{v,l})$ is the hard decision of the received LLR vector $\alpha_{v,l}$. $\mathcal{C}_{v,l,1}$ is obtained by flipping the $k_{M,l}$-th bit of $\mathcal{C}_{v,l,0}$, where $k_{M,l}$ is the index of the LLR element with the smallest absolute value among $\alpha_{v,l}$.


Each decoding path splits into two paths and has two associated candidate codewords. Alg.~\ref{algo: CG} calculates $2L$ expanded path metrics PM$_l^j$ for $l=0,1,\cdots,L-1$ and $j=0,1$ to select $L$ codewords passed to the parent node. The min$_L$ function in Alg.~\ref{algo: CG} finds the $L$ smallest values among $2L$ input expanded path metrics. Once $\beta_{v,l}$ for $l=0,1,\cdots,L-1$ are computed, decoding path $a_l$ is copied to decoding path $l$ before further operations.

\begin{algorithm}
\DontPrintSemicolon
\label{algo: CG}
\SetKwInOut{Input}{input}\SetKwInOut{Output}{output}

\Input{$\alpha_{v,0}, \alpha_{v,1},\cdots,\alpha_{v,L-1}$}
\Output{$\beta_{v,0}, \beta_{v,1},\cdots,\beta_{v,L-1}$; $a_0,a_1,\cdots,a_{L-1}$}
\BlankLine

\For{$l=0$ \KwTo $L-1$} {
$k_{M,l} = \underset{k\in\{0,1,\cdots,I_v-1\}}{\arg\!\min}|\alpha_{v,l}[k]|$\;
$\mbox{NM}_l^0 = 0$; $\mathcal{C}_{v,l,0} = h(\alpha_{v,l})$\;
$\mbox{NM}_l^1 = |\alpha_{v,l}[k_{M,l}]|$; $\mathcal{C}_{v,l,1} = \mbox{Flip}(\mathcal{C}_{v,l,0}, k_{M,l})$\;
$\mbox{PM}_l^j = \mbox{PM}_l + \mbox{NM}_l^j$ for $j=0,1$\;
}
$(\mbox{PM}_{a_0}^{b_0}, \cdots,\mbox{PM}_{a_{L-1}}^{b_{L-1}}) = \min_L(\mbox{PM}_0^0, \mbox{PM}_0^1,\cdots,\mbox{PM}_{L-1}^1)$
\For{$l=0$ \KwTo $L-1$} {
$\beta_{v,l} = \mathcal{C}_{v,a_l,b_l}$; $\mbox{PM}_l = \mbox{PM}_{a_l}^{b_l}$\;
}
\caption{The proposed CG algorithm}
\end{algorithm}



\subsubsection{Proposed MBS Algorithm} \label{ssec: mld}


%

When an FP node is activated, each current decoding path expands to $2^{I_v}$ paths, each of which is associated with a candidate codeword. Similar to the CG algorithm, the proposed MBS algorithm calculates $L$ codewords passed to the parent node and $L$ path indices, $a_0, a_1,\cdots,a_{L-1}$. The calculation of returned codewords are shown as follows.
\begin{itemize}
\item For each candidate codeword $\mathcal{C}_{v,l}^j$,  calculate its corresponding node metric NM$_l^j$ for $j=0,1,\cdots,2^{I_v}-1$ and $l=0,1,\cdots,L-1$.
\item Calculate $2^{I_v}L$ expanded path metrics PM$_l^{j}$ for $l=0,1,\cdots,L-1$ and $j=0,1,\cdots,2^{I_v}-1$.
\item Find $L$ expanded path metrics among $2^{I_v}L$ ones. The correspondent candidate codewords are passed to the parent node $v_p$.
\end{itemize}
To calculate the node metric, we propose a new method with low computational complexity. In the literature, two methods can be used: the direct-mapping method (DMM) shown in Eq.~(\ref{equ: direct_method}) and the recursive channel
combination (RCC)~\cite{chenrong_tsp}. In terms of computational
complexity, the former needs $2^{I_v}(2^{n-t}-1)L$ additions,
where $N=2^n$ and $t$ is the layer index
of an FP node $v$. The RCC needs $(\sum_{i=1}^{n-t-1}2^i2^{2^{n-t-i}}+2^{I_v})L$ additions.
Compared to the DMM, the RCC approach needs fewer additions. For our RLLD algorithm, we want to compute these $2^{I_v}$ node metrics in parallel. However, the parallel hardware implementations of the DMM and RCC algorithms require large area consumption. This will be discussed in more detail in Section~\ref{ssec: slmld_archi}.

In this paper, a hardware efficient node metric computation method, which takes advantage of both the
DMM and the RCC, is proposed. The proposed method, referred to as the DR-Hybrid (DRH) method, is shown in
Alg.~\ref{algo: dr_hyp}, where $\mathcal{C}_{v,l}^j[2i:2i+1] =
(\mathcal{C}_{v,l}^j[2i],\mathcal{C}_{v,l}^j[2i+1])$, and $r$ $(0\leqslant r \leqslant 3)$ is represented by a binary tuple of length two, i.e. $r = r_0+2r_1$. In
our method, the RCC approach is used to calculate $\theta_{l,i}$ first. Then,
the DMM is carried out.

\begin{algorithm}
\caption{DR-Hybrid method}
\label{algo: dr_hyp}
\LinesNumbered
\For{$l=0$ \KwTo $L-1$}{
\tcc{----------RCC----------------}
\For{$i=0$ \KwTo $2^{n-t-1}-1$}{
\For{$r=0$ \KwTo $3$}{
$\theta_{l,i}[(r_0, r_1)] = (1-2r_0)\alpha_{v,l}[2i]+(1-2r_1)\alpha_{v,l}[2i+1];$
}
}
\tcc{----------DMM----------------}
\For{$j=0$ \KwTo $2^{I_v}-1$}{
$\mbox{NM}_{l}^j=\sum_{i=0}^{2^{n-t-1}-1}\theta_{l,i}[\mathcal{C}_{v,l}^j[2i:2i+1]].$
}
}
\end{algorithm}

The DRH method needs $4\times 2^{n-t-1}+2^{I_v}(2^{n-t-1}-1)$ additions. Take $X_0=8$ and $X_1=16$ as an example, the DMM, RCC and DRH methods need 3840, 864 and 1824 additions. Though our DRH method needs more additions than the RCC, it results in a more area efficient hardware implementation when all $2^{I_v}$ node metrics are computed in parallel, since the RCC method needs more complex multiplexors.


Once we have $2^{I_v}L$ node metrics and corresponding candidate codewords, $2^{I_v}L$ expanded path metrics PM$_l^j$ = PM$_l$ + NM$_l^j$ for $l=0,1,\cdots,L-1$ and $j=0,1,\cdots,2^{I_v}-1$ can be computed. The next step is selecting $L$ returned codewords and their corresponding expanded path metrics.

Since directly finding the $L$ minimum values from $2^{I_v}L$ ones is computationally intensive and lacks efficient hardware implementations, a bitonic sequence based sorter~\cite{jun_low_mem_list} (BBS) with $2^{I_v}L$ inputs is able to fulfill this task. Such a BBS takes $2^{I_v-1}L(\sum_{i=1}^{s-1}i)+2^{I_v-2}L$ compare-and-switch (CS) units~\cite{jun_low_mem_list}, where each of them has one comparator and two 2-to-1 multiplexors and $s=\log_2(2^{I_v}L)$.
In order to simplify the hardware implementation, a two-stage sorting scheme was proposed in~\cite{chenrong_tsp}, where the first stage selects $q\mbox{ }(q<L)$ smallest node metrics from $2^{I_v}$ ones for each decoding path. The second stage selects the $L$ smallest metrics from the $Lq$ expanded path metrics produced by the first stage. Compared with the direct sorting scheme~\cite{jun_low_mem_list, yuan_low_latency}, the hardware implementation of the two-stage sorting scheme is more efficient at the cost of certain error performance degradation.

In this paper, our MBS algorithm employs the two-stage sorting scheme and improves the first stage in the following two aspects:
\begin{itemize}
\item Instead of using a fixed $q$, our MBS algorithm employs a dynamic $q_{I_v,L} (q_{I_v,L}\leqslant L)$, which is a power of 2 and depends on both $I_v$ and $L$.
\item An approximated sorting (ASort) method, which leads to an efficient hardware implementation, is used to select $q_{I_v,L}$ metrics from $2^{I_v}$ ones, though these sorted metrics are not always the $q_{I_v,L}$ smallest ones.
\end{itemize}
Our ASort method is illustrated as follows:
\begin{itemize}
\item When $2^{I_v}\leqslant 2L$, the BBS with $2L$ inputs and $L$ outputs is used to select the $q_{I_v,L}$ minimum node metrics from $2^{I_v}$ ones.
\item When $2^{I_v} > 2L$, all $2^{I_v}$ node metrics are divided into $q_{I_v,L}$ groups:
\end{itemize}
\begin{equation}\nonumber
\underbrace{\mbox{NM}_l^0, \cdots, \mbox{NM}_l^{m-1}}_{\mbox{group 1}},\cdots,\underbrace{\mbox{NM}_l^{(q_{I_v,L}-1)m}, \cdots, \mbox{NM}_l^{q_{I_v,L}m-1}}_{\mbox{group } q_{I_v,L}}.
\end{equation}
Here $m= \frac{2^{I_v}}{q_{I_v,L}}$. The two minimum node metrics of each group are first computed. The BBS computes the minimum $q_{I_v,L}$ node metrics among $2q_{I_v,L}$ ones.

After the first stage of sorting, the number of expanded path metrics $N_e$ could be $2L, 4L,\cdots, L\times L$. The second stage of sorting is the same as that in~\cite{chenrong_tsp}. A binary tree of $2L$-$L$ BBSs are employed to sort the final $L$ minimum expanded path metrics. Take $N_e=4L$ as an example, there are $4L$ extended path metrics: PM$_{l_0}^{j_0}$, PM$_{l_1}^{j_1}$, $\cdots$, PM$_{l_{4L-1}}^{j_{4L-1}}$, then PM$_{l_0}^{j_0}$, $\cdots$, PM$_{l_{2L-1}}^{j_{2L-1}}$ and PM$_{l_{2L}}^{j_{2L}}$, $\cdots$, PM$_{l_{4L-1}}^{j_{4L-1}}$ are applied to two 2$L$-$L$ BBSs, respectively. Thus, $2L$ metrics are selected. Then the 2$L$-$L$ BBS is employed again to generated the final $L$ minimum extended path metrics: PM$_{l'_0}^{j'_0}$, PM$_{l'_1}^{j'_1}$, $\cdots$, PM$_{l'_{L-1}}^{j'_{L-1}}$.

\subsection{Parameters of Our RLLD Algorithm} \label{sec: dicus_parameter}
For our RLLD algorithm, the returned codewords from rate-1 nodes with $I_v> X_{th}$ are obtained by making hard decisions on the received LLR vectors. The other rate-1 nodes are processed by our CG algorithm. Note that both the hard decision approach and our CG algorithm could cause potential error performance degradation since ideally we should consider $2^{I_v}$ candidate codewords for each decoding path. With more rate-1 nodes (decreasing $X_{th}$) being processed by the hard decision approach, the decoding latency could be reduced at the cost of more error performance degradation. Besides, in order to save computations, path metrics remain unchanged when a rate-0 node is activated, which may cause error performance degradation.

The choices of $X_0$ and $X_1$ are tradeoffs between implementation complexity and achieved decoding latency reduction. Ideally, we want $X_0$ and $X_1$ to be as large as possible so that more data bits could be decoded in parallel. Since the number of adders needed by Alg.~\ref{algo: dr_hyp} is proportional to $2^{X_0}X_1$, the values of $X_0$ and $X_1$ are limited by hardware implementations.

For the two-step sorting scheme of our MBS algorithm, we want $q_{I_v,L}$ to be as small as possible so that the sorting complexity could be minimized. However, reducing $q_{I_v,L}$ could degenerate the resulting error performance, since ideally we need to consider the $L$ most reliable candidate codewords for each decoding path. As a result, the selections of $q_{I_v,L}$ are tradeoffs between sorting complexity and error performance.


\subsection{Comparison with Related Algorithms} \label{sec: algo_cmp}
If we perform the SC based list decoding algorithms~\cite{ido_it, tree_list_dec} on a tree, then all $2N-1$ nodes of the tree will be activated. For our RLLD algorithm, denote $n_a$ as the number of activated nodes. Then we have $n_a < 2N-1$, where $n_a$ is determined by the block length $N$, the code rate, the locations of frozen bits and the parameters $X_0$ and $X_1$. $X_0$ and $X_1$ are used to identify all FP nodes. The reduction of the number of activated nodes will transfer into reduced decoding latency and increased throughput. Take the (8, 3) polar code in Fig.~\ref{fig: dec_tree} as an example, suppose $X_0=1$ and $X_1 = 2$, then only 5 nodes (nodes 0, 1, 2, 5, and 6) need to be activated by our RLLD algorithm, whereas the algorithms in~\cite{ido_it, tree_list_dec} need to activate all 15 nodes.



The CA-SCL decoding algorithm was also performed on a binary tree in~\cite{gabi_low_latency}. Compared with the low-latency list decoding algorithm~\cite{gabi_low_latency}, our RLLD algorithm employs the proposed MBS algorithm to process FP nodes, while FP nodes were processed by activating its child nodes in~\cite{gabi_low_latency}. Our MBS algorithm results in decreased decoding latency at the cost of potential error performance loss. Besides, our RLLD algorithm takes a simpler approach when a rate-1 node is activated.
When a rate-1 node is activated, a Chase-like algorithm was used to calculate the $L$ codewords passed to the parent node in~\cite{gabi_low_latency}. Compared to the Chase-like algorithm, our CG algorithm has lower computational complexity and is more suitable for hardware implementation because:

(1) The Chase-like algorithm in~\cite{gabi_low_latency} was performed over log-likelihoods (LL) domain while our method is performed over LLR domain. Compared with our LLR based method, it takes more additions to calculate related metrics for the Chase-like algorithm.

(2) For each decoding path, the Chase-like algorithm considers $1+{c \choose 1} + {c \choose 2}$ candidate constituent codewords, where $c=2$ in~\cite{gabi_low_latency}. In contrast, our method considers only two constituent codewords, which leads to simpler hardware implementations.

(3) In order to find the $L$ best decoding paths and their constituent codewords, the Chase-like algorithm creates a candidate path list. The final $L$ candidates are determined by inserting and removing elements from the list. The Chase-like algorithm is suitable for software implementations. However, the hardware implementations of the Chase-like algorithm has not been discussed in~\cite{gabi_low_latency}. On the other hand, with a bitonic based sorter~\cite{jun_low_mem_list} (BBS), the $L$ most reliable decoding paths can be decided in parallel for our CG algorithm.

\subsection{Simulation Results} \label{ssec: sim1}
\begin{figure} [hbt]
\centering
  \includegraphics[width=3in]{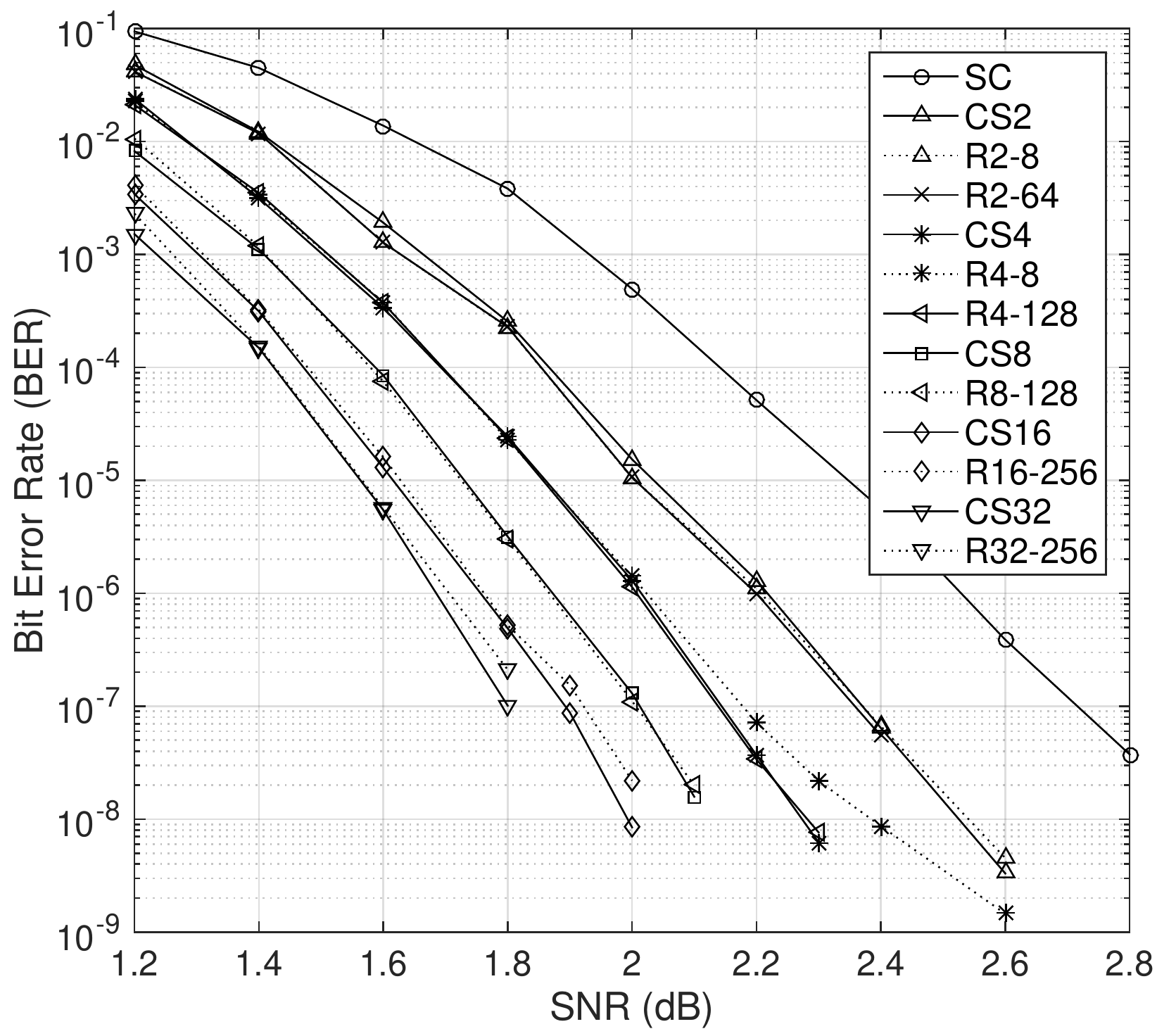}
  \caption{BER performance for an (8192, 4096) polar code}\label{fig: ber8192}
\end{figure}

For an (8192, 4096) polar code, the bit error rate (BER) performances of the proposed RLLD algorithm as well as other algorithms are shown in Fig.~\ref{fig: ber8192}. In Fig.~\ref{fig: ber8192}, CS$x$ denotes the CA-SCL decoding algorithm with $L=x$, where CRC-32 is used. R$x$-$y$ denotes our RLLD algorithm with $L=x$ and $X_{th}=y$. The values of $q_{I_v, L}$'s under different list sizes and $I_v$'s are shown in Table~\ref{tab: m_i}. For all simulated algorithms, the additive white Gaussian noise (AWGN) channel and binary phase-shift keying (BPSK) modulation are used. For all simulated RLLD algorithms, $X_0=8$ and $X_1=16$.

\begin{table}[hbt]
  \centering
  \caption{The Values of $q_{I_v, L}$'s under Different List Sizes and $I_v$'s}
  \label{tab: m_i}
  \footnotesize
  \begin{tabular}{c|c||c|c|c|c|c|c|c|c}
    \hline
   & $I_v$& 1 & 2 & 3 & 4 & 5 & 6 & 7& 8 \\ \hline\hline
   \multirow{5}{*}{$L$}& 2   & 2         & 2        & 2        & 2         & 2         & 2        &  2       & 2 \\
       &4& 2         & 4        & 4        & 4        & 4         & 4        &  4       & 2 \\
      &8 & 2         &4         & 8        &8         &8          &8         & 4        & 2 \\
      &16 & 2         &4         & 8        &8         &8          &8         & 8        & 2 \\
      & 32& 2         &4         & 8        &8         &8          &4         & 4        & 2 \\ \hline
  \end{tabular}
\end{table}

Based on the simulation results shown in Fig.~\ref{fig: ber8192}, we observe that R2-8 performs nearly the same as CS2 and R2-64. When the list size increases, compared with CS4, R4-8 shows obvious error performance degradation when BER is below $10^{-7}$. The degradation is reduced by increasing $X_{th}$ to 128, as we observe that R4-128 performs nearly the same as CS4. When the list size further increases (e.g. $L=16$ and 32), at low BER level, the error performance degradation exists even when $X_{th} =256$. As shown in Fig.~\ref{fig: ber8192}, R16-256 and R32-256 are worse than CS16 and CS32 when BER is below $10^{-5}$ and $10^{-6}$, respectively. Note that for the (8192, 4096) polar code in this paper, $I_v$ of a rate-1 node is at most 256. The simulation results of a (1024, 512) polar code show similar phenomena.

Depending on the specific list size, it seems that our RLLD algorithm has performance degradation compared to the CA-SCL algorithm at certain BER values even when all rate-1 nodes are processed by the proposed CG algorithm.
There are several reasons for the error performance degradation:

(1) For our RLLD algorithm, when a rate-1 node with $I_v \leqslant X_{th}$ is activated, only the two most reliable constituent codewords are kept. When list size $L$ is large, there may not be enough candidate codewords to include the correct codeword, since our CG algorithm could miss certain good candidate codewords.

(2) When a rate-1 node with $I_v> X_{th}$ is activated, only the most reliable candidate codeword is considered for each decoding path, which could also cause error performance degradation.

(3) During the first sorting stage of our MBS algorithm, when $2^{I_v} \geqslant L$, $q_{I_v,L}$ is selected to be no greater than $L$ for certain $I_v$ values for efficient hardware implementation. As a result, we may lose certain good candidate codewords due to the limitation on $q_{I_v,L}$.


\section{High Throughput List Polar Decoder Architecture} \label{sec: llldec}
\subsection{Top Decoder Architecture} \label{ssec: top_archi}

\begin{figure} [hbt]
\centering
  \includegraphics[width=3.0in]{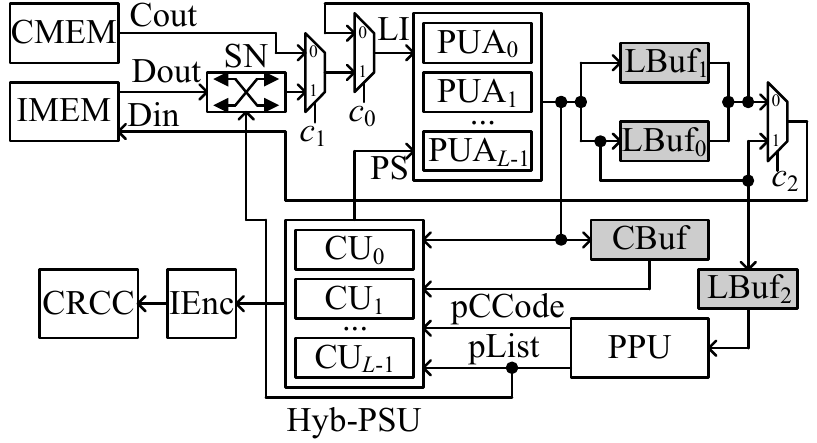}
  \caption{Decoder top architecture}\label{fig: dec_top}
\end{figure}

In this paper, based on the proposed RLLD algorithm, a high throughput list decoder architecture, shown in Fig.~\ref{fig: dec_top}, for polar codes is proposed. In Fig.~\ref{fig: dec_top}, the channel message memory (CMEM) stores the received channel LLRs, and the internal LLR message memory (IMEM) stores the LLRs generated during the SC computation process. With the concatenation and split method in our prior work~\cite{jun_low_mem_list}, the IMEM is implemented with area efficient memories, such as register file (RF) or SRAM. The proposed architecture has $L$ groups of processing unit arrays (PUAs), each of which contains $T$ processing units~\cite{gross_polar1} (PUs) and is capable of performing either the $f$ or the $g$ computation in Eqs.~(\ref{equ: f_comp_simplified}) and~(\ref{equ: g_comp}), respectively. The hybrid partial sum unit (Hyb-PSU) in Fig.~\ref{fig: dec_top} consists of $L$ computation units, CU$_0$, CU$_1$, $\cdots$, CU$_{L-1}$, which are responsible for updating the partial sums of $L$ decoding paths, respectively. The path pruning unit (PPU) in Fig.~\ref{fig: dec_top} finds the list indices and corresponding constituent codewords for $L$ survival decoding paths, respectively. The control of our decoder architecture can be designed based on the instruction RAM based methodology in~\cite{fast_polar_SC_gross}.

Both our high throughput list decoder architecture in Fig.~\ref{fig: dec_top} and that in~\cite{jun_low_mem_list} employ a partial parallel processing method. Besides, both architectures contain a channel message memory and internal message memory. However, compared to the architecture in~\cite{jun_low_mem_list}, the major improvements of our list decoder architecture are:

(a) Instead of LL messages, our high throughput list decoder architecture employs LLR messages, which result in more area efficient internal and channel message memories.

(b) The PPU in Fig.~\ref{fig: dec_top} implements our CG and MBS algorithms, while the PPU in~\cite{jun_low_mem_list} is just a sorter which selects $L$ values among $2L$ ones. Due to the proposed PPU, our decoder architecture achieves much higher throughput than that in~\cite{jun_low_mem_list}.

(c) Our list decoder architecture employs a novel Hyb-PSU, which is more area and energy efficient than that in~\cite{jun_low_mem_list}. Our Hyb-PSU is based on the proposed index based partial sum computation algorithm. When a decoding path needs to be copied to another one, instead of copying partial sums directly our Hyb-PSU copies only decoding path indices. In contrast, the PSU in~\cite{jun_low_mem_list} copies path sums directly, which incurs additional energy consumption. Our Hyb-PSU stores most of the partial sums in area efficient memories, while the PSU in~\cite{jun_low_mem_list} stores all the partial sums in area demanding registers. Hence, our Hyb-PSU is scalable for larger block lengths.

\subsection{Memory Efficient Quantization Scheme}
\label{sec: quan}
For an SC or SCL decoder, the message memory occupies a large part of the overall decoder area~\cite{gross_polar1, jun_low_mem_list}. An SCL decoder needs a channel message memory and an internal message memory. For an LLR based SCL decoder, the channel memory stores $N$ channel LLR messages. The internal message memory stores $Ln$ LLR matrices: $P_{l,t}$ for $l=0,1,\cdots,L-1$ and $t=1,2,\cdots,n$, where $P_{l,t}$ has $2^{n-t}$ LLR messages.

For a fixed point implementation of our RLLD algorithm, it is straightforward to quantize all LLRs in the internal memory with $Q$ bits.
In this paper, a memory efficient quantization (MEQ) scheme is proposed to reduce the size of the internal memory. $f(a,b)$ in Eq.~(\ref{equ: f_comp_simplified}) has the same magnitude range as those of $a$ and $b$, while the magnitude range of $g(a,b,s)$ in Eq.~(\ref{equ: g_comp}) is at most twice of those of $a$ and $b$ ($s$ is either 0 or 1). Since $P_{0,t},P_{1,t},\cdots,P_{L-1,t}$ are computed based on $P_{0,t-1},P_{1,t-1},\cdots,P_{L-1,t-1}$, for a decoding path $l$, the LLRs in $P_{l,t_1}$ may need a greater magnitude range than that of the LLRs in $P_{l,t_2}$, where $t_1 > t_2$. Suppose each channel LLR is quantized with $Q_c$ bits, the proposed MEQ scheme is as follows:

(1) Suppose all LLRs within the internal memory are quantized with $Q_m$ bits, determine the minimal $Q_m$ such that the error performance degradation of the fixed point performance is negligible.

(2) Let $t_1, t_2,\cdots,t_r$ be $r$ integers, where $t_1\leqslant t_2\leqslant \cdots \leqslant t_{r}\leqslant n$ and $r=Q_m-Q_c$. Denote $\mathbf{P}_t = (P_{0,t},P_{1,t}\cdots,P_{L-1,t})$. Suppose LLRs associated with $\mathbf{P}_1,\mathbf{P}_2,\cdots,\mathbf{P}_{t_1}$ are quantized with $Q_c$ bits and the remaining LLRs are quantized with $Q_m$ bits. Decide the maximal $t_1$ such that the resulting fixed point error performance degradation is negligible. Once $t_1$ is decided, suppose the LLRs within $\mathbf{P}_{t_1+1}, \mathbf{P}_{t_1+2},\cdots, \mathbf{P}_{t_2}$ are all quantized with $Q_c+1$ bits, find the maximal $t_2$ such that the corresponding error performance degradation is negligible. In this way, $t_3,\cdots,t_r$ are decided in a serial manner so that $\mathbf{P}_{t_i+1}, \mathbf{P}_{t_i+2},\cdots, \mathbf{P}_{t_{i+1}}$ are quantized with $Q_c+i$ bits for $1\leqslant i \leqslant r-1$, and $\mathbf{P}_j$ are quantized to $Q_m$ bits for $j> t_r$.

With the proposed MEQ scheme, the number of bits saved for the internal memory is
\begin{equation}
N_B = \sum_{j=1}^{r+1}\sum_{t=t_{j-1}+1}^{t_j}L2^{n-t}(Q_c+j-1),
\end{equation}
where $t_0= 0$ and $t_{r+1} = n$ are introduced for convenience.

\begin{figure} [hbt]
\centering
  \includegraphics[width=2.4in]{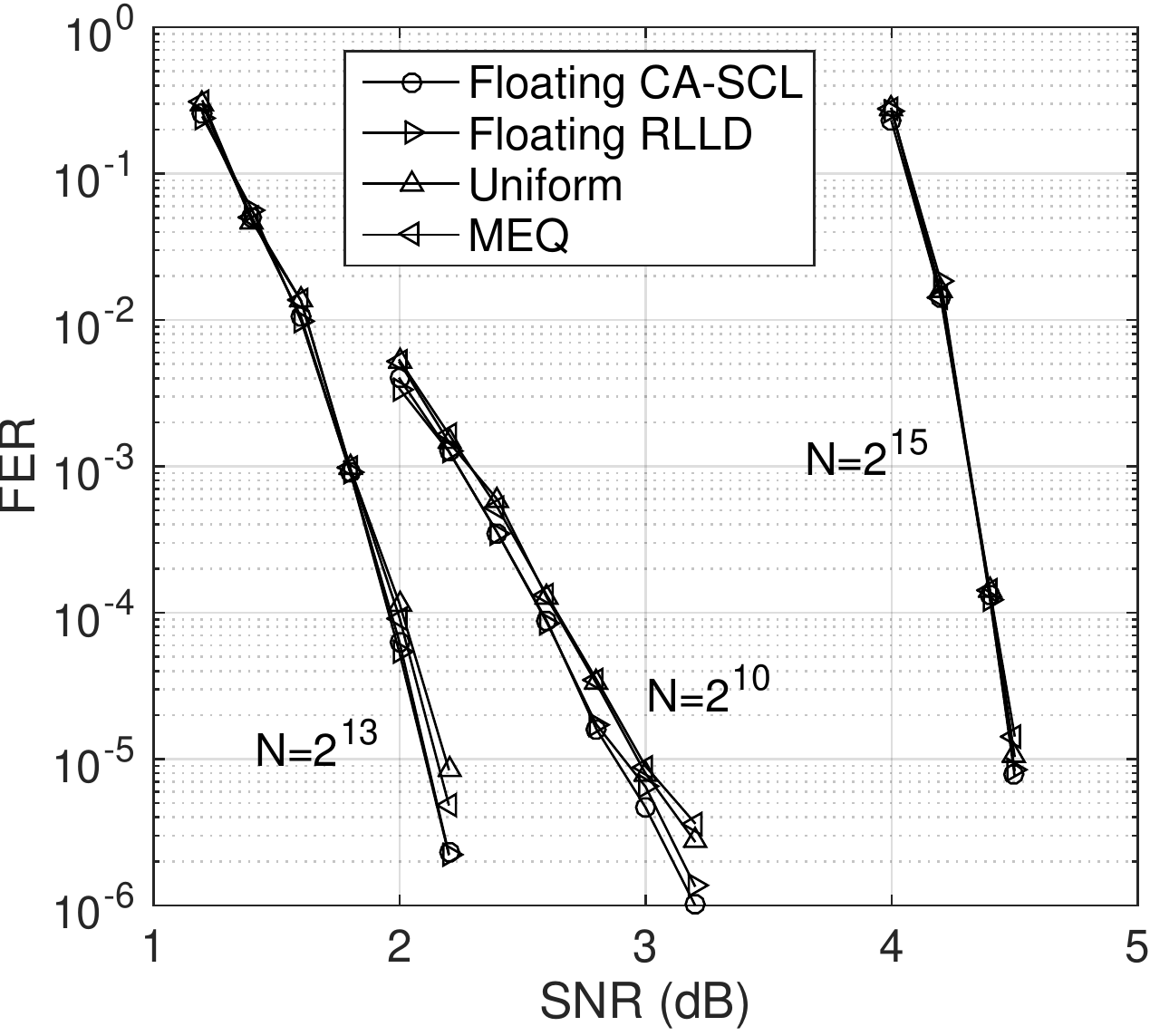}
  \caption{Effects of the proposed MEQ scheme on the error performances}\label{fig: fer_quan}
\end{figure}

In order to show the effectiveness of our MEQ scheme, the error performances of our RLLD algorithm with the proposed MEQ scheme are shown in Fig.~\ref{fig: fer_quan}, where the RLLD algorithm with our MEQ scheme is compared with the floating-point CA-SCL decoding algorithm, floating-point RLLD algorithm, and RLLD algorithm with a uniform quantization scheme for three different polar codes, (1024, 512), (8192, 4096) and (32768, 29504) with $X_{th}=32, 128, 1024$, respectively. For all fixed-point decoders, each channel LLR is quantized with $Q_c=5$ bits. For the RLLD algorithm with uniform quantization, each LLR in the internal memory is quantized with $Q_m=6$ bits for the length $2^{10}$ and $2^{13}$ polar codes. For the polar code with a length of $2^{15}$, the uniform quantization takes 7 bits. For our MEQ scheme, $Q_m=7$. Since $Q_m-Q_c =2$, we need to determine two integers, $r_1$ and $r_2$, for our MEQ scheme. When $N=2^{10}$, $2^{13}$ and $2^{15}$, $(r_1, r_2)$ = (1,2), (3,4) and (4,5), respectively. As shown in Fig.~\ref{fig: fer_quan}, the performance degradation caused by our MEQ scheme is small. Compared with the uniform quantization, the proposed MEQ scheme reduces the number of stored bits by 4.5\%, 13.5\% and 27.2\% for $N=2^{10}$, $2^{13}$ and $2^{15}$, respectively. For all the simulation results shown in Fig.~\ref{fig: fer_quan}, list size $L=4$.

\subsection{Proposed path pruning unit} \label{ssec: slmld_archi}

When a rate-1 node with $I_v\leqslant X_{th}$ or an FP node is activated, each decoding path splits into multiple ones and only the $L$ most reliable paths are kept. The PPU in Fig.~\ref{fig: dec_top} implements our CG and MBS algorithms, and is responsible for calculating $L$ returned codewords, $\beta_{v,0}, \beta_{v,1},\cdots,\beta_{v,L-1},$ and $L$ path indices, $a_0, a_1,\cdots,a_{L-1}$. For $l=0,1,\cdots,L-1$, decoding path $l$ copies from decoding path $a_l$ before further decoding steps.

\begin{figure} [hbt]
\centering
  \includegraphics[width=2.8in]{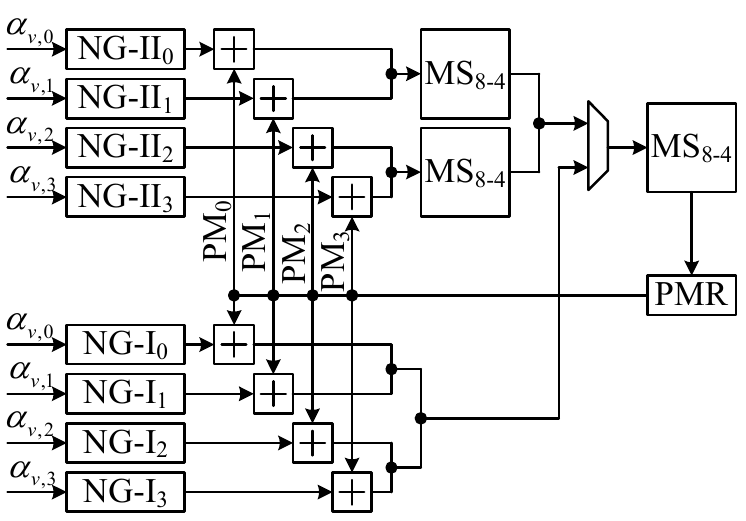}
  \caption{The proposed architecture for PPU}\label{fig: slmld}
\end{figure}

Take $L=4$ as an example, the proposed PPU is shown in Fig.~\ref{fig: slmld}, which can be easily adapted to other $L$ values. Our PPU in Fig.~\ref{fig: slmld} has two types of node metric generation (NG) units, NG-I and NG-II, which compute the node metrics for a rate-1 node and an FP node, respectively. NG-I$_l$ and NG-II$_l$ correspond to decoding path $l$.
For decoding path $l$, the expanded path metrics PM$_l^j$'s are obtained by adding the node metrics to the path metric PM$_l$, which is stored in the path metric registers (PMR) and initialized with 0.

When a rate-1 node is activated, NG-I$_l$ outputs two node metrics for $l=0,1,\cdots,L-1$. After $2L$ expanded path metrics are computed, a stage of metric sorter (MS$_{2L-L}$) selects the $L$ minimum metrics and their corresponding codewords from $2L$ ones. The metrics sorter MS$_{2L-L}$ implements the min$_L$ function in Alg.~\ref{algo: CG} and can be constructed with a BBS. When an FP node is activated, $L$ NG-II modules implement the first part of our two-stage sorting scheme. For each decoding path, $q_{I_v,L}$ node metrics and their correspondent codewords are computed. The tree of metric sorters sort the $L$ minimum metrics among $q_{I_v,L}L$ ones. This is achieved by $\log_2 q_{I_v,L}$ stages of metric sorters, where $q_{I_v,L}$ is a power of 2. The output expanded path metrics of the last stage of metric sorter are saved in the PMR. The corresponding codewords of the selected $L$ expanded path metrics are also chosen. The related circuitry is omitted for brevity.

\begin{figure} [hbt]
  \centering
  \includegraphics[width=2.3in]{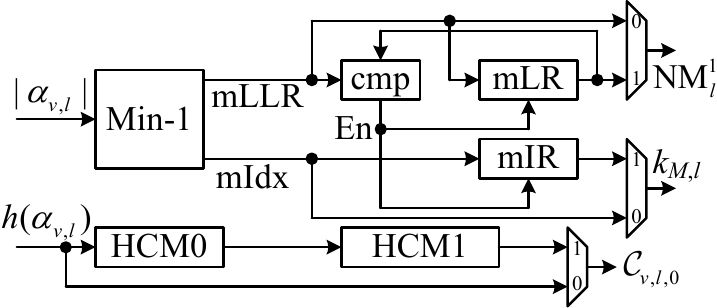}
  \caption{Hardware architecture of the proposed NG-I$_l$}\label{fig: ngi}
\end{figure}

The micro architecture of NG-I$_l$ is shown in Fig.~\ref{fig: ngi}. The most complex part of NG-I$_l$ is finding the minimum LLR magnitude and its corresponding index among the LLR vector $|\alpha_{v,l}| \triangleq (|\alpha_{v,l}[0]|, |\alpha_{v,l}[1]|, \cdots, |\alpha_{v,l}[I_v-1]|)$. Since the node metric of the most reliable candidate codeword is always 0, we need to compute $\mbox{NM}_l^1=|\alpha_{v,l}[k_{M,l}]|$ in Fig.~\ref{fig: ngi}, which is the node metric of the second most reliable candidate codeword, with a corresponding index $k_{M,l}$. For our list decoder architecture, for each decoding path, at most $T$ LLRs are computed in one clock cycle, since we have only $T$ PUs per decoding path. The Min-1 unit in Fig.~\ref{fig: ngi} is capable of finding the minimum value, mLLR, and its corresponding index, mIdx, from at most $T$ parallel inputs. When $I_v \leqslant T$, $\mbox{NM}_l^1$ = mLLR and $k_{M,l}$ = mIdx. $\mathcal{C}_{v,l,0}$ = $h(\alpha_{v,l})$ in Fig.~\ref{fig: ngi} is the hard decision of $\alpha_{v,l}$, which is the most reliable candidate codeword. The second most reliable candidate codeword is obtained by flipping the $k_{M,l}$-th bit of $\mathcal{C}_{v,l,0}$.

When $I_v > T$, suppose $T$ is a power of 2, then $I_v$ can be divided by $T$. During each clock cycle, only $T$ LLRs are fed to NG-I$_l$, and the minimum value and its corresponding index are computed in a partial parallel way. The minimum value and associated index of the first $T$ inputs are stored in mLR and mIR, respectively. The minimum value of the second group of $T$ inputs is compared with the current value stored in mLR, and is stored in mLR if it is smaller than the current value of mLR. This repeats until the whole LLR vector $\alpha_{v,l}$ is processed. At last, the minimum value of $|\alpha_{v,l}|$ and its index are stored in mLR and mIR, respectively. The hard decoding of $\alpha_{v,l}$ is stored in the hard decoded constituent codeword memory (HCM0), and is copied to HCM1 when the second most reliable constituent codeword is computed.

\begin{figure}[htbp]
\centering
\includegraphics[width=2.8in]{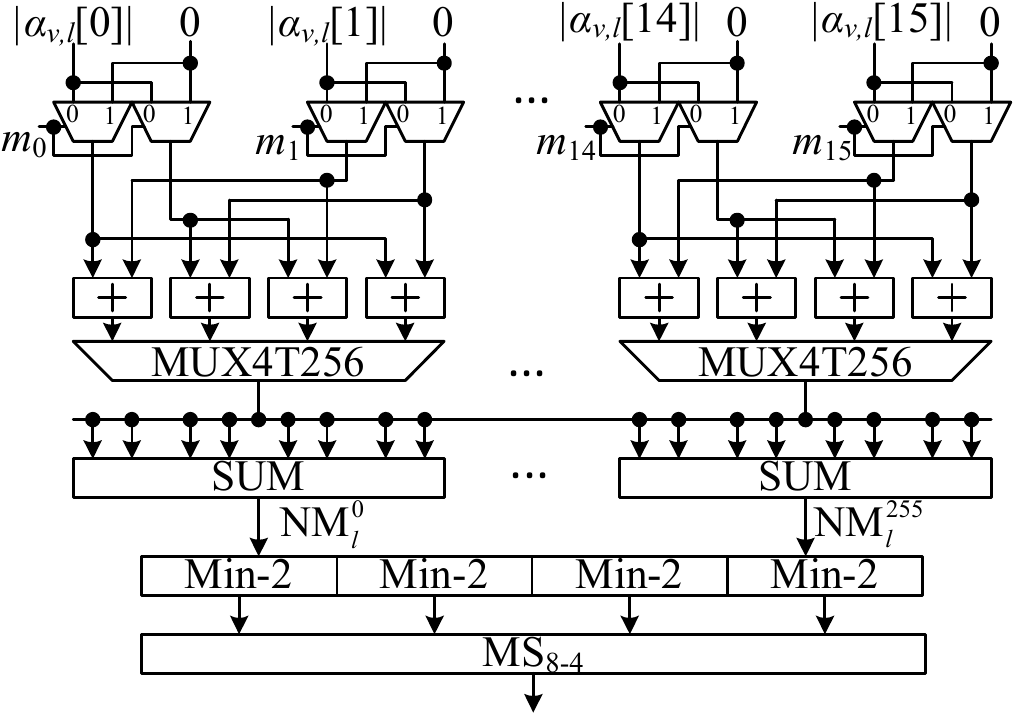}
\caption{Architecture of NG-II$_l$}\label{fig:ML16}
\end{figure}

The micro-architecture of NG-II$_l$ under $X_0=8$ and $X_1=16$ is shown in Fig.~\ref{fig:ML16}, where the block MUX4T256 includes 256 4-to-1 multiplexers. Our NG-II$_l$ consists of two parts: the first part calculates $2^{I_v}$ node metrics, NM$_l^0$, NM$_l^1$, $\cdots$, NM$_l^{2^{I_v}-1}$, based on Alg.~\ref{algo: dr_hyp}, and the second part implements the first stage sorting of our MBS algorithm. For $L=4$, when $2^{I_v} > 2L$, the $2^{I_v}$ metrics are first divided into four groups. The Min-2~\cite{min2} block is modified slightly to find the two minimum node metrics and their associated indices for each metric group. The MS$_{8-4}$ block calculates the final output metrics. When $2^{I_v} = 2L = 8$, the MS$_{8-4}$ blocks work directly on the $2L = 8$ expanded path metrics. When $2^{I_v} \leqslant L$, the expanded path metrics are output directly. As shown in Figs.~\ref{fig: slmld} to~\ref{fig:ML16}, our PPU has long critical path delay, since there are many levels of logic from the inputs to outputs. Pipelines should be used to improve overall decoder frequency.

Based on the DMM method in Eq.~(\ref{equ: direct_method}), the node metric computation part needs $2^{I_v}(2^{n-t}-1)L$ adders and $2^{I_v}2^{n-t}L$ 2-to-1 multiplexers, where $N=2^n$ and $t$ is the layer index of an FP node $v$. Based on the RCC method, it takes $(\sum_{i=1}^{n-t-1}2^i2^{2^{n-t-i}}+2^{I_v})L$ adders, $2^{I_v+1}L$ $2^{2^{n-t-1}}$-to-1 multiplexers and $4\times 2^{n-t-1}L$ 2-to-1multiplexers.
In contrast, based on our DRH method, it takes $4\times 2^{n-t-1}+2^{I_v}(2^{n-t-1}-1)$ adders, $2^{I_v}2^{n-t-1}$ 4-to-1 and $4\times 2^{n-t-1}$ 2-to-1 multiplexers. Table~\ref{tab:ML_16} compares hardware resources needed by the DMM, RCC and DR-Hybrid methods when $X_0=8$, $X_1=16$, and $\alpha_{v,l}[j]\text{ }(0 \leq j < 2^{n-t})$ is a 6-bit LLR. As shown in Table~\ref{tab:ML_16}, the DRH method requires the smallest total area. Besides, the implementations based on DMM, RCC and DRH have roughly the same critical path delay.

\begin{table}[htbp]
\begin{center}
\caption{Hardware resources needed by different methods per list}
\label{tab:ML_16}
\begin{tabular}{|c|c|c|c|}
\hline
& DMM & RCC & DRH \\ \hline
\# of adders & 3840 & 864 & 1824 \\ \hline
\# of MUX$_{2-1}$ & 4096 & 32 & 32 \\ \hline
\# of MUX$_{4-1}$ & 0 & 0 & 2048 \\ \hline
\# of MUX$_{256-1}$ & 0 & 512 & 0 \\ \hline
total area (\# of NANDs) & 313,967 & 1,673,810 & 229,449 \\ \hline
\end{tabular}
\end{center}
\end{table}

\subsection{Proposed hybrid partial sum unit} \label{ssec: top_archi}
For the list decoder architectures in~\cite{tree_list_dec, jun_low_mem_list}, all partial sums are stored in registers and the partial sums of decoding path $l'$ are copied to decoding path $l$ when decoding path $l'$ needs to be copied to decoding path $l$. The PSU in~\cite{tree_list_dec} and~\cite{jun_low_mem_list} needs $L(N-1)$ and $L(\frac{N}{2}-1)$ single bit registers to store all partial sums, respectively. Thus, for large $N$, the register based PSU architectures in~\cite{tree_list_dec, jun_low_mem_list} are inefficient for two reasons. First, the area of the PSU is linearly proportional to $N$. For large $N$ (e.g. $N>2^{15}$), the area of PSU is large since registers are usually area demanding. Second, the power dissipation due to the copying of partial sums between different decoding paths is high when $N$ is large. 

\subsubsection{Proposed Index Based Partial Sum Computation Algorithm} \label{ssec: mld}
In order to avoid copying partial sums directly, an index based partial sum computation (IPC) algorithm is proposed in Algorithm~\ref{algo: psum_comp}, where $p_l[z]\mbox{ }(l=0,1,\cdots,L-1\mbox{ and }z=0,1,\cdots,n)$ is a list index reference. $C_{l,z}$ for $l=0,1,\cdots,L-1$ and $z=0,1,\cdots,n$ are partial sum matrices~\cite{ido_it, jun_low_mem_list}. $C_{l,z}$ has $2^{n-z}$ elements, each of which stores two binary bits.

For our RLLD algorithm, once a rate-0, rate-1 or an FP node sends $L$ codewords to its parent node, the partial sum computation is performed after decoding path pruning. Let $t$ denote the layer index of such a node $v$. Let $(B_{n-1}, B_{n-2},\cdots, B_0)$ denote the binary representation of the index of the last leaf node belonging to node $v$, where $B_{n-1}$ is the most significant bit. Let $t_e = n-j$, where $j$ is the smallest integer such that $B_j =0$. If $B_j=1$ for $j=0,1,\cdots,n-1$, $t_e=0$.
Once $\beta_{v,0}, \beta_{v,1},\cdots,\beta_{v,L-1}$ are calculated, decoding path $l'$ may need to be copied to path $l$ before the following partial sum computation. Under this circumstance, the index references are first copied, where $p_{l'}[z]$ is copied to $p_l[z]$ for $z=t, t-1, \cdots,0$. The lazy copy algorithm was proposed in~\cite{ido_it} to avoid copying partial sums directly. However, the lazy copy algorithm is not suitable for hardware implementation due to complex index computation. The PSU in~\cite{jun_low_mem_list} copies all partial sums belonging one decoding path to the corresponding locations of another decoding path.

\begin{algorithm}
\DontPrintSemicolon
\label{algo: psum_comp}
\SetKwInOut{Input}{input}\SetKwInOut{Output}{output}

\Input{$t_e, t, (\beta_{v,0}, \beta_{v,1},\cdots,\beta_{v,L-1})$}
\Output{$C_{l,t_e}[j][0]$ for $l=0,1,\cdots,L-1$ and $j=0,1,\cdots,2^{n-t_e}$}
\BlankLine

\For{$l=0$ \KwTo $L-1$} {
\For{$j=0$ \KwTo $2^{n-t}-1$} {
\If{$v$ is the left child node of its parent node} {
$C_{l,t}[j][0] = \beta_{v,l}[j]$; $p_l[t] = l$\;
}\lElse {
$C_{l,t}[j][1] = \beta_{v,l}[j]$
}
}
}
\lIf{$v$ is the left child node of its parent node} {
\textbf{exit}
}
\For{$l=0$ \KwTo $L-1$} {
\For{$z = t-1$ \KwTo $t_e$} {
\For{$j=0$ \KwTo $2^{n-z-1}$} {
$v_0 = C_{p_l[z+1],z+1}[j][0]$; $v_1= C_{l,z+1}[j][1]$\;
\If{$z == t_e$}{
$C_{l,z}[2j][0] = v_0\oplus v_1$; $C_{l,z}[2j+1][0] = v_1$\;
$p_l[z+1] = p_l[z]=l$\;
}\Else{
$C_{l,z}[2j][1] = v_0\oplus v_1$; $C_{l,z}[2j+1][1] = v_1$\;
$p_l[z+1] = l$\;
}
}
}
}
\caption{Index Based Partial Sum Computation (IPC) Algorithm}
\end{algorithm}



\begin{figure*} [hbt]
\centering
  \includegraphics[width=5.6in]{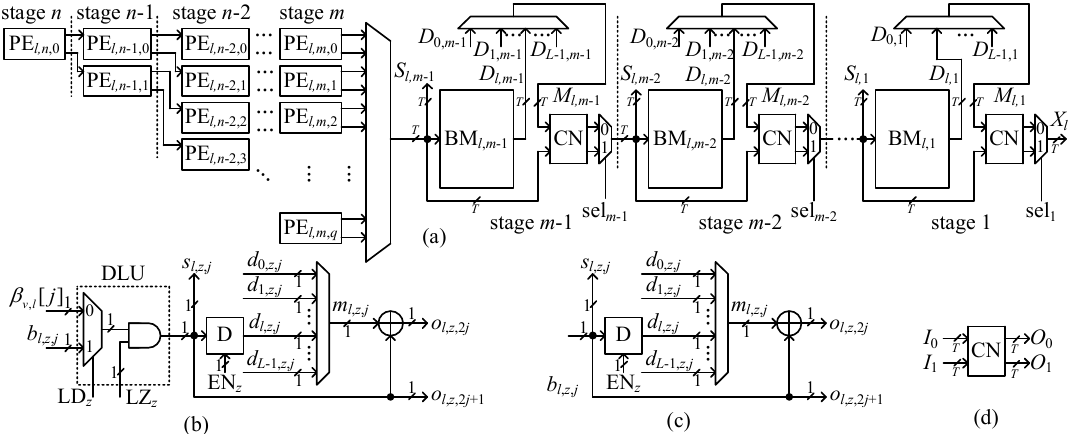}
  \caption{(a) Top architecture of CU$_l$. (b) Type-I PE. (c) Type-II PE. (d) Inputs and outputs of the CN.}\label{fig: psu_top}
\end{figure*}

\subsubsection{Micro Architecture of the Proposed Hybrid Partial Sum Unit}
Based on our IPC algorithm, a Hyb-PSU is proposed with two improvements. First, some partial sums are stored in memory, while others are stored in registers. Second, instead of partial sums, only list index matrices are copied. These two improvements reduce the area and power overhead of partial sum computation unit when $N$ is large. The Hyb-PSU consists of $L$ computation units, CU$_0$, CU$_1$, $\cdots$, CU$_{L-1}$, where the micro architecture of CU$_l$ is shown in Fig.~\ref{fig: psu_top}(a) and is described as follows.

(a) Let $m$ be a predefined integer parameter. For block length $N=2^n$, CU$_l$ consists of $n$ stages, where the first $n-m+1$ stages are a binary tree of the type-I and type-II unit processing elements (PEs) shown in Figs.~\ref{fig: psu_top}(b) and~\ref{fig: psu_top}(c), respectively. Stage $z$ ($z\geqslant m$) has $2^{n-z}$ PEs. Each of the remaining $m-1$ stages has the same circuitry.

(b) Two types of PEs are used in the PE tree in Fig.~\ref{fig: psu_top}(a). Suppose the maximal length of a constituent codeword that is returned from a rate-0, rate-1 or FP node is $2^\mu$, then stage $z$ ($z \geqslant n- \mu$) employs only the type-I PEs. The remaining stages in the PE tree employ the type-II PEs.

(c) Compared with the type-II PE, the type-I PE has an extra data load unit (DLU). For PE$_{l,z,j}$ within stage $z$ ($j=0,1,\cdots,2^{n-z}-1$), the binary outputs, $o_{l,z,2j}$ and $o_{l,z,2j+1}$, are connected to $b_{l,z-1, 2j}$ and $b_{l,z-1, 2j+1}$, respectively. The wired connections are not shown in Fig.~\ref{fig: psu_top}(a) for simplicity.

(d) BM$_{l,z}$ ($z \leqslant m-1$) is a bit memory with $c_{w,z} =\frac{2^{n-z}}{T}$ words, where each word contains $T$ bits. $T$ is the number of processing elements belonging to a decoding path in a partial parallel list decoder. For our memory compiler, if $c_{w,z}$ is greater than a threshold value, then BM$_{l,z}$ is implemented with an RF. If $c_{w,z}$ is even greater than another threshold value, then BM$_{l,z}$ is implemented with an SRAM.

(e) The connector module (CN) has two $T$-bit inputs and two $T$-bit outputs. The connections between the outputs and inputs are
\begin{eqnarray}
\left\{\begin{array}{llll}
O_0[2j]& = & I_0[j] \oplus I_1[j] & 0\leqslant j < T/2\\
O_0[2j+1] & = & I_1[j] & 0\leqslant j < T/2 \\
O_1[2j-T] & = & I_0[j]  \oplus I_1[j] & T/2\leqslant j < T\\
O_1[2j+1-T] & = & I_1[j] & T/2\leqslant j < T
\end{array}
\right.\label{eqn: xor_eq}
\end{eqnarray}

(f) For our Hyb-PSU, $L$ computation units are needed. For each PE within CU$_l$, $m_{l,z,j}$ in Figs.~\ref{fig: psu_top}(b) and~\ref{fig: psu_top}(c) is the output of an $L$-to-1 multiplexer whose inputs are $d_{0,z,j}$, $d_{1,z,j}$, $\cdots$, $d_{L-1,z,j}$, where $L-1$ of them are from other computation units. For each CN, $M_{l,z}$ is the output of an $L$-to-1 multiplexer whose inputs are $D_{0,z}, D_{1,z}, \cdots, D_{L-1,z}$. 

\subsubsection{Computation Schedule of Our Hybrid Partial Sum Unit}
Once the returned $L$ codewords $\beta_{v,0}, \beta_{v,1},\cdots,\beta_{v,L-1}$ are computed, the path pruning unit also outputs $L$ indices $a_0, a_1,\cdots,a_{L-1}$, where $a_l$ needs to be copied to decoding path $l$. For $l=0,1,\cdots,L-1$, $\beta_{v,l}$ is first loaded into stage $t$ by the DLU in Fig.~\ref{fig: psu_top}(b), and the output partial sums in Alg.~\ref{algo: psum_comp} come out from stage $t_e$. For stage $t$, if $\beta_{v,l}$ is sent from a rate-0 node, then the control signal LZ$_t$ is 0, since $\beta_{v,l}$ is a zero vector.  Otherwise, LD$_t$ = 0 and LZ$_t$ = 1. For the other stages, LD$_z$ = 1 and LZ$_z$ = 1 ($z\neq t$).

For all partial sums within the partial sum matrix $C_{l,z}$, we divide them into two sets: $\mathbf{C}_{l,z}^0$ and $\mathbf{C}_{l,z}^1$, where $\mathbf{C}_{l,z}^0$ consists of $C_{l,z}[j][0]$ for $j=0,1,\cdots,2^{n-z}-1$ and $\mathbf{C}_{l,z}^1$ consists of the other partial sums within $C_{l,z}$. For each $C_{l,z}$, our Hyb-PSU stores only $\mathbf{C}_{l,z}^0$ in the registers or bit memory of stage $z$.  As shown in Alg.~\ref{algo: psum_comp}, for $z=t-1$ to $t_e+1$, $\mathbf{C}_{l,z}^1$ is computed in serial. At last, $\mathbf{C}_{l,t_e}^0$ is computed. For our Hyb-PSU, after loading the returned $L$ codewords into stage $t$, for $z=t-1$ to $t_e+1$, $\mathbf{C}_{l,z}^1$ is computed on-the-fly and passed to the next stage as shown in Fig.~\ref{fig: psu_top}.

When $t_e\geqslant m$, $\mathbf{C}_{l,t_e}^0$ is computed in one clock cycle and is output from stage $t_e$, where $C_{l,t_e}[j][0]$ is set to $s_{l,t_e,j}$ produced by the type-I and type-II PEs for $j=0,1,\cdots,2^{n-t_e}-1$. When $t_e<m$, $\mathbf{C}_{l,t_e}^0$ is computed in $2^{n-t_e}/T$ cycles, and $T$ updated partial sums are computed in each clock cycles. Since decoding path $a_l$ needs to be copied to path $l$, for $z=t,t-1,\cdots, t_e+1$, the computation of $\mathbf{C}_{l,z}^1$ is based on $\mathbf{C}_{a_l,z+1}^0$ and $\mathbf{C}_{l,z+1}^1$. Hence, the multiplexers within stage $z$ are configured so that $m_{l,z,j} = d_{a_l,z,j}$ for $z\geqslant m$. When $z<m$, $M_{l,z} = Q_{a_l,z}$.

\subsubsection{Comparisons with Related Works}
Compared to the partial sum computation architectures in~\cite{tree_list_dec, jun_low_mem_list}, the proposed Hyb-PSU architecture has advantages in the following two aspects.

(1) The proposed Hyb-PSU is a scalable architecture. The PSU architectures in~\cite{tree_list_dec, jun_low_mem_list} require $L(N-1)$ and $L(N/2-1)$ single bit registers, where $N=2^n$ is the block length. Hence, they will suffer from excessive area overhead when the block length $N$ is large. In contrast, the proposed Hyb-PSU stores $L(N-1)$ bits and most of these bits are stored in RFs or SRAMs, which are more area efficient than registers. 

(2) The architectures in~\cite{tree_list_dec, jun_low_mem_list} copies partial sums of a decoding path to another decoding path when needed, while our Hyb-PSU copies only index references. We define the copying of a single bit from one register to another as a single copy operation. When decoding path $l'$ needs to be copied to path $l$, the PSU in~\cite{jun_low_mem_list} requires $N_1 = 2^{n-1}-1$ copy operations, while our Hyb-PSU needs only $N_2 = (n+1)\log_2 L$ copy operations. Since the value of $L$ for practical hardware implementation is small, our lazy copy needs much fewer copy operations than direct copy.

In this paper, when $L=4$ and $T=128$, for $N=2^{13}$ and $2^{15}$, the proposed hybrid partial sum unit architecture is implemented with $m=3$ and $m=5$, respectively, under a TSMC 90nm CMOS technology. Our partial sum computation unit consumes an area of 0.779mm$^{2}$ and 1.31mm$^2$ for $N=2^{13}$ and $N=2^{15}$, respectively.

To the best of our knowledge, those decoder architectures in~\cite{tree_list_dec, jun_low_mem_list, chuan_list, yuan_low_latency} are the only for SC based list decoding algorithms of polar codes. However, in~\cite{tree_list_dec,chuan_list, yuan_low_latency}, the partial sum computation unit architecture was not discussed in detail and the implementation results on the PSU alone are not shown. Hence, we compare our proposed Hyb-PSU with that in~\cite{jun_low_mem_list}.
When $L=4$, the partial sum unit architecture in~\cite{jun_low_mem_list} for $N=2^{13}$ and $2^{15}$  consumes an area of 1.011mm$^{2}$ and 3.63mm$^2$, respectively, under the same CMOS technology. All PSUs are synthesized under a frequency of 500MHz. Our Hyb-PSU achieves an area saving of 23\% and 63\% for block length $2^{13}$ and $2^{15}$, respectively.


\subsection{Latency and Throughput} \label{ssec: dec_cycle}

For the proposed high throughput decoder architecture, the number of clock cycles, $N_D$, used on the decoding of a codeword depends on the block length,  the code rate and the positions of frozen bits.
For our RLLD algorithm, let $N_V$ be the number of nodes (except the root node) visited in $G_n$. Let $S_V$ denote the set of indices of visited nodes (except the root node). Let $S_V'$ be a subset of $S_V$ and $S_V'$ consists of rate-1 nodes with $I_v \leqslant X_{th}$ and all FP nodes. For $v_i \in S_V$, let $t_i$ be the layer index of node $v_i$ for $i=0,1,\cdots,N_V-1$.
Then
\begin{equation}\label{equ: cycle_cnt}
N_D = \sum_{i=0}^{N_V-1}(N_L^{(i)}+N_P^{(i)})+N_C,
\end{equation}
where $N_L^{(i)} = \lceil\frac{2^{n-t_i}}{T}\rceil$ is the number of clock cycles needed to calculate the LLR vectors sending to node $v_i$. $N_P^{(i)}$ is the number of clock cycles used by our PPU when $v_i$ is activated. Note that decoding path splits only if node $v_i$ is a rate-1 node with $I_v\leqslant X_{th}$ or an FP node. Hence, $N_P^{(i)} = 0$ if $v_i \not\in S_V'$. 
If $v_i\in S_V'$, $N_P^{(i)}\neq 0$ and depends on the node type, $X_{th}$, $q_{I_v,L}$, $T$, $L$ and the number of pipeline stages in our PPU. This will be discussed in more detail in Section~\ref{sec: imp_results}.

Since our list decoder outputs $x_0^{N-1}$ instead of $u_0^{N-1}$, we need to obtain $u_0^{N-1}$ based on $x_0^{N-1}$ before calculating the CRC checksum of the information bits. A partial-parallel polar encoder~\cite{polar_encoder} can be used and the corresponding latency is $N/T$ when $T$ bits are fed to the encoder in parallel. For the computation of CRC, a partial parallel CRC unit~\cite{crc_lanman2015} can be used, and the corresponding latency is also $N/T$. As a result, $N_C=\frac{2N}{T}$ is the number of clock cycles due to encoding and CRC checksum computation.

The latency of our decoder is $T_L = N_D/f$, where $f$ is the decoder frequency. Since we are using CRC for output final data word, we calculate the net information throughput (NIT) of our decoder, where $\mbox{NIT}=\frac{(N\times R-h)f}{N_D-N_C}$, where $h$ is the CRC checksum length. Here, the latency due to the CRC checksum computation does not affect out decoder throughput, since our decoder can work on the next frame once our Hyb-PSU begins to output decoded codewords for the current frame.

\begin{table*}[hbt]
  \centering
  \caption{Implementation Results for $N=2^{10}, R=0.5$}
  \label{tab:imp_result_n_10}
  \begin{threeparttable}
  \footnotesize
  \begin{tabular}{c||c|c|c||c|c|c||c|c|c||c|c|c|c||c}
    \hline
     &       \multicolumn{3}{c||}{proposed} & \multicolumn{3}{c||}{\cite{llr_list_tsp}} & \multicolumn{3}{c||}{\cite{jun_low_mem_list}\ddag}&  \multicolumn{4}{c||}{\cite{yuan_low_latency}}&\cite{chenrong_tsp}\\ \hline\hline
     $L$                           & 2   &4    &8          & 2&4&8 & 2&4&8 & \multicolumn{2}{c|}{2} & \multicolumn{2}{c||}{4} & 4\\ \hline
     Frequency (MHz)         &423 &403 &289      &847  &794  &637  & 507&492&462 & 500$^*$ &361\dag &400$^*$ & 288\dag & 500\\ \hline
     Cell Area (mm$^2$)    &1.98 &3.83 &7.22 &0.88 &1.78 &3.85 & 1.23&2.46&5.28 &1.06$^*$ &2.03\dag&2.14$^*$ & 4.10\dag & 1.403\\ \hline
     \# of Decoding Cycles   &337  &371 &404 &2592 &2649 &2649 & 2592&2592&3104 &  \multicolumn{2}{c|}{1022} &  \multicolumn{2}{c||}{1022}& 1290\\ \hline
     NIT (Mbps)                 &666 &570&374   &168  &154 &123    &93     &91   &71 & 250$^*$ & 180\dag&200$^*$ &144\dag& 186\\ \hline
     Latency (us)               &0.79&0.92 &1.39  & 3.06&3.34&4.16 & 5.11 &5.26 &6.72 & 2.04$^*$& 2.83\dag&2.55$^*$ &3.54\dag&2.58\\ \hline
     AE (Mbps/mm$^2$)    &336 &148&51  & 191 &86 &32            &76    &37 &13 & 237$^*$ & 88\dag &94$^*$ &35\dag&132\\ \hline
  \end{tabular}
    \begin{tablenotes}[para,flushleft]
    \ddag The decoder architecture in~\cite{jun_low_mem_list} has been re-synthesized under the TSMC 90nm CMOS technology. $^*$ These are the original implementation results based on a 65nm CMOS technology. \dag These are the scaled results under the TSMC 90nm CMOS technology.
  \end{tablenotes}
  \end{threeparttable}
\end{table*}

\begin{table*}[hbt]
  \centering
  \caption{Implementation Results for $N=2^{13}, R=0.5$}
  \label{tab:imp_result_n_13}
  \begin{threeparttable}
  \footnotesize
  \begin{tabular}{c||c|c|c||c|c|c||c|c|c||c}
    \hline
     &       \multicolumn{3}{c||}{proposed} & \multicolumn{3}{c||}{\cite{llr_list_tsp}\dag} & \multicolumn{3}{c||}{\cite{jun_low_mem_list}\ddag}&\cite{chenrong_tsp}\ddag\\ \hline\hline
     $L$                            & 2    &4    &8         &2      &4     &8            &2    &4    &8 & 4\\ \hline
      Frequency (MHz)         &416 &398 &289      &847  &794  &637         &467 &434 &434 & 434\\ \hline
     Cell Area (mm$^2$)    &3.42 &6.46 &12.26  &6.48 &12.73 &28.04     & 3.97 &7.93 &17.45 &7.02\\ \hline
     \# of Decoding Cycles   &2146 &2367 &2576  &20736 &20736 &20736 & 20736 &20736 &24832&11488 \\ \hline
     NIT (Mbps)                 &839  &723 &479     &167  &156 &125   &92  &85 &71& 153\\ \hline
     Latency (us)               &5.16 &5.94 &8.91   & 24.48&26.11&32.55     & 44.40 &47.78 &58.56&26.47 \\ \hline
     AE (Mbps/mm$^2$)     &245 &111 &39       & 26 &12 &4.6               &23    &11 &4.1 &21.79 \\ \hline
  \end{tabular}
    \begin{tablenotes}[para,flushleft]
    \dag These results are estimated conservatively.
\ddag The decoder architectures in~\cite{jun_low_mem_list, chenrong_tsp} have been re-synthesized under the TSMC 90nm CMOS technology. The number of PU per decoding path is 128.
  \end{tablenotes}
  \end{threeparttable}
\end{table*}

\begin{table*}[hbt]
  \centering
  \caption{Implementation Results for $N=2^{15}, R=0.9004$}
  \label{tab:imp_result_n_15}
  \begin{threeparttable}
  \footnotesize
  \begin{tabular}{c||c|c|c||c|c|c||c|c|c||c}
    \hline
     &       \multicolumn{3}{c||}{proposed} & \multicolumn{3}{c||}{\cite{llr_list_tsp}\dag} & \multicolumn{3}{c||}{\cite{jun_low_mem_list}\ddag}&\cite{chenrong_tsp}\ddag\\ \hline\hline
     $L$                           & 2&4&8                   &2     &4     &8             &2     &4   &8 &4\\ \hline
      Frequency (MHz)         &367 &359 &286       &847  &794  &637         & 398&389 &389 &389\\ \hline
     Cell Area (mm$^2$)    &6.22 &11.89 &23.1   &25.68 &50.41 &111.08  & 8.59&17.54 &34 &15.5\\ \hline
     \# of Decoding Cycles   &6070  &6492 &6895  &96576 &96576 &96576  &96576 &96576 &126080&63606 \\ \hline
     NIT (Mbps)                 &1949 &1772 &1323   &258  &242 &194            &121  &118 &90& 180\\ \hline
     Latency (us)               &16.53 &18.08 &24.11  & 114.02&121.63&151.61&242.65 &248.26 &324.1 &163.5\\ \hline
     AE (Mbps/mm$^2$)    &313 &149&57             & 11 &4.8 &1.75               &14    &6.72 &2.64 & 11.61\\ \hline
  \end{tabular}
    \begin{tablenotes}[para,flushleft]
    \dag These results are estimated conservatively.
\ddag The decoder architectures in~\cite{jun_low_mem_list, chenrong_tsp} have been re-synthesized under the TSMC 90nm CMOS technology. The number of PU per decoding path is 128.
  \end{tablenotes}
  \end{threeparttable}
\end{table*}

\section{Implementation Results and Comparisons} \label{sec: imp_results}
To compare with prior works, we implement our high throughput list decoder architecture for three polar codes with lengths of $2^{10}$, $2^{13}$ and $2^{15}$, respectively, and rates 0.5, 0.5 and 0.9, respectively. The last polar code is intended for storage applications. For each code, three different list sizes are considered: $L=2,4,8$. All our decoders are synthesized under the TSMC 90nm CMOS technology using the Cadence RTL compiler. The area efficiency (AE) of a partly parallel decoder architecture depends on the number of PUs. In order to make a fair comparison with prior works in~\cite{llr_list_tsp,jun_low_mem_list,chenrong_tsp}, the number of PUs for each decoding path of our implemented decoders is selected to be 64 when $N=2^{10}$. When $N=2^{13}$ and $2^{15}$, the number of PUs per decoding path is 128 for our decoders. The list decoders in~\cite{yuan_polar} are based on a line architecture, which requires $\frac{N}{2}$ PUs.

A total of 3, 4 and 6 pipeline stages, respectively, are inserted in the PPU for decoders with $L=2$, 4 and 8, respectively. The number of pipeline stages needed for our PPU is determined by the longest data path. For each $v_i\in S_V'$, if node $v_i$ is a rate-1 node with $I_v \leqslant X_{th}$, $N_P^{(i)}$ depends on the number of PUs in a decoding path: when $I_v \leqslant T$, $N_P^{(i)}=2$ for all our implemented decoders; otherwise, $N_P^{(i)}=4$ for all our decoders, since the minimum value of a received LLR vector is calculated in a partial parallel way, which incurs extra clock cycles. When node $v_i$ is an FP node, $N_P^{(i)}$ relates to $q_{I_v,L}$. Depending on the detailed value of $q_{I_v,L}$, we may use different data paths when computing the $L$ minimum expanded path metrics. The locations of all pipelines are arranged so that fewer clock cycles are needed when the $q_{I_v,L}$ is smaller. In Table~\ref{tab: addi_cycles}, we list the detailed value of $N_P^{(i)}$ with respect to $I_v$ and $L$.


The selection of $X_{th}$ is a trade-off between AE and error performance. When increasing $X_{th}$, more rate-1 nodes will be processed by our CG algorithm. Hence, $N_D$ increases and the resulting NIT decreases. Meanwhile, the corresponding error performance is better especially in high SNR region. Our high throughput list decoder architecture supports all $X_{th}$ values. For all our implemented decoders, $X_{th}$ is large enough so that all rate-1 nodes are processed by our CG algorithm. In this setup, for each implemented decoder, $N_D$ is maximized with respect to $X_{th}$, and hence the throughput of our decoder architecture in Tables~\ref{tab:imp_result_n_10}, \ref{tab:imp_result_n_13} and \ref{tab:imp_result_n_15} is the \emph{minimum} achieved by our decoders. For each code, the corresponding error performance is better than that of the RLLD with the MEQ in Fig.~\ref{fig: fer_quan}.

\begin{table}[hbt]
  \centering
  \caption{$N_P^{(i)}$ with Respect to $I_v$ and $L$}
  \label{tab: addi_cycles}
  \footnotesize
  \begin{tabular}{c||c|c|c|c|c|c|c|c}
    \hline
   $I_v$& 1 & 2 & 3 & 4 & 5 & 6 & 7& 8 \\ \hline\hline

  $L=2$    & 2         & 2        & 3        & 3        & 3        & 3        &  3       & 3 \\ \hline
$L=4$    & 2         & 4        & 4        & 4        & 4         & 4        &  4       & 3 \\ \hline
  $L=8$   & 2         &3         & 4        &5         &5          &6        & 5        & 3 \\ \hline
  \end{tabular}
\end{table}

The implementation results are shown in Table~\ref{tab:imp_result_n_10},~\ref{tab:imp_result_n_13} and~\ref{tab:imp_result_n_15}. The implementation results show that our decoders outperform existing SCL decoders~\cite{llr_list_tsp,jun_low_mem_list,yuan_low_latency} in both decoding latency and area efficiency. Compared with the decoders of~\cite{llr_list_tsp}, the area efficiency and decoding latency of our decoders are 1.59 to 32.5 times and 3.4 to 6.8 times better, respectively. The area efficiency and decoding latency of our decoders are 3.9 to 21.5 times and 5.5 to 13 times better, respectively, than the decoders of~\cite{jun_low_mem_list}. Compared with decoders of~\cite{chenrong_tsp}, our decoders improve the area efficiency and decoding latency by 1.12 to 12 times and 2.8 to 9 times, respectively. When $N=2^{10}$, the area efficiency and decoding latency of our decoders are 3.8 to 4.2 times and 3.58 to 3.84 times better, respectively, than the decoders of~\cite{yuan_low_latency}. Compared with the decoders of~\cite{yuan_low_latency}, our decoders would show more significant improvements in area efficiency and decoding latency when $N$ is larger.


Based on the implementation results shown in Tables~\ref{tab:imp_result_n_10},~\ref{tab:imp_result_n_13} and~\ref{tab:imp_result_n_15}, it is observed that when the block length is fixed, as the list size $L$ increases, the area efficiency and decoding latency will decrease and increase, respectively, because:
\begin{itemize}
\item It takes more memory to store internal LLRs when $L$ increases.
\item The number of pipeline stages within our PPU will increase when $L$ increases, which in turn increases the overall decoding clock cycles.
\end{itemize}

The latency reduction and area efficiency improvement of our decoders are due to the reduced number of nodes activated in the decoding. However, the area and frequency overhead of the proposed PPU somewhat dilute the effects due to decoding clock cycles reduction. For example, our decoder reduces the number of decoding cycles to approximately $\frac{1}{7}$ of that of the decoders in~\cite{llr_list_tsp} for $L=2$, 4 and 8. However, the reduction in decoding cycles does not fully transfer into the improvement in decoding latency and area efficiency. Based on our implementation results, take $L=2$ as an example, the PPU occupies 61.99\%, 40.16\% and 25.40\% of the area of the whole decoder, for $N=2^{10}$, $2^{13}$ and $2^{15}$, respectively. Compared with the decoders with $N=2^{10}$ and $2^{13}$, the effects on the area efficiency caused by the area overhead of PPU are smaller for decoders with $N=2^{15}$. Keeping $T$ unchanged, as $N$ increases, the area of the PPU increases very slowly while the total area of all LLR memories is proportional to $N$. Hence, for larger $N$, PPU occupies a smaller percentage of the total area of a whole decoder. When list size $L$ is fixed, as $N$ increases, the latency reduction and area efficiency improvement compared with other decoders in the literature will be greater.


\section{Conclusion} \label{sec: conclusion}
In this paper, a reduced latency list decoding algorithm is proposed for polar codes. The proposed list decoding algorithm results in a high throughput list decoder architecture for polar codes. A memory efficient quantization method is also proposed to reduce the size of message memories. The proposed list decoder architecture can be adapted to large block lengths due to our hybrid partial sum unit, which is area efficient. The implementation results of our high throughput list decoder demonstrate significant advantages over current state-of-the-art SCL decoders.

\bibliographystyle{IEEEbib}
\bibliography{refs_latest}

\end{document}